\documentclass[aps,pre,twocolumn,amsmath,amssymb,showpacs]{revtex4-1}
\usepackage{graphicx}% Include figure files
\usepackage{subfigure}% Include subfigure 
\usepackage{dcolumn}% Align table columns on decimal point
\usepackage{bm}% bold math
\usepackage{color}

\usepackage{gensymb}

\begin{document}

\title{Vertical dynamics of a horizontally-oscillating active object in a 2D granular medium}

%\titlerunning{}

\author{Ling Huang}
 \email{walter250034@yahoo.com}
%\author{Raphael Blumenfeld}%
\affiliation{%
College of Science, National University of Defense Technology, Changsha 410073, China
%% This line break forced with \textbackslash\textbackslash
}%
\author{Xianwen Ran}
 \email{ranxianwen@163.com}
%\author{Raphael Blumenfeld}%
\affiliation{%
College of Science, National University of Defense Technology, Changsha 410073, China
%% This line break forced with \textbackslash\textbackslash
}%

\author{Raphael Blumenfeld}%
 \email{rbb11@cam.ac.uk}
\affiliation{
College of Science, National University of Defense Technology, Changsha 410073, China,
}%
\affiliation{
Imperial College London, London SW7 2AZ, UK,
}%
\affiliation{
Cavendish Laboratory, Cambridge University, JJ Thomson Avenue, Cambridge CB3 0HE, UK
}%

% \collaboration{CLEO Collaboration}%\noaffiliation

\date{\today}% It is always \today, today,

\begin{abstract}
We use a DEM simulation and analytical considerations to study the dynamics of a self-energised object, modelled as a disc, oscillating horizontally within a two-dimensional bed of denser and smaller particles. 
We find that, for given material parameters, the immersed object (IO) may rise, sink or not change depth, depending on the oscillation amplitude and frequency, as well as on the initial depth. With time, the IO settles at a specific depth that depends on the oscillation parameters. We construct a phase diagram of this behaviour in the oscillation frequency and velocity amplitude variable space.
We explain the observed rich behaviour by two competing effects: climbing on particles, which fill voids opening under the disc, and sinking due to bed fluidisation. 
We present a cavity model that allows us to derive analytically general results, which agree very well with the observations and explain quantitatively the phase diagram. 
Our specific analytical results are the following.
(i) Derivation of a critical frequency, $f_c$, above which the IO cannot float up against gravity. We show that this frequency depends only on the gravitational acceleration and the IO size.
(ii) Derivation of a minimal amplitude, $A_{min}$, below which the IO cannot rise even if the frequency is below $f_c$. We show that this amplitude also depends only on the gravitational acceleration and the IO size.
(iii) Derivation of a critical value, $g_c$, of the IO's acceleration amplitude, below which the IO cannot sink. We show that the value of $g_c$ depends on the characteristics of both the IO and the granular bed, as well as on the initial IO's depth.

\end{abstract}

%\keywords{Brazil nut effect \and Granular dynamics}

\maketitle

\section{\label{Intro}Introduction}

The organisation dynamics of granular matter (GM) is key to understanding many essential industrial processes, such as sorting, stirring \cite{Soetal06,Guetal13,Guetal14,Guetal15} and transport \cite{Knetal96,Esetal10}. It is also the basis for modelling segregation and pattern formation in GM comprising particles of different characteristics \cite{ShMu98,Moetal01,Sh04,Meetal96}. 
Due to the complexity of granular dynamics, it is useful to reduce the problem, in the first instance, to the dynamics of a large object immersed in a medium of smaller particles, driven by a simple periodic oscillation. 
Such dynamics are relevant to a number of applications and have been studied in several contexts: impact cratering and penetration of objects into GM \cite{WoWh98,Lo04a,Lo04b,KaDu07,Cletal15}, the Brazil nut effect \cite{ShMu98,Moetal01,Sh04}, uplifting of pipes\cite{Chetal08}, and animal locomotion in sand \cite{Do96,Maetal09,Maetal11a}. 

A range of models has been proposed in these contexts, in particular for the rise of objects driven horizontally and `plowing' through GM \cite{Coetal11,Peetal11}, as well as for the lift forces on such immersed objects (IOs) \cite{Soetal06,Guetal14,Waetal03,Chetal03,Dietal11,Maetal11b,PoDi13}. Numerical and experimental observations report a complicated dependence of these forces on the driving velocities and the depth of the IO. Consequently, there is currently no general theory for this phenomenon and existing models are mainly phenomenological. 
In particular, it is unclear what determines whether a large IO would rise or sink under different types of driving processes. In comparison to the development of understanding of conventional fluids, where rise is dominated straightforwardly by the density difference between the IO and the fluid, the current state of modelling of this phenomenon in granular fluids is pre-Archimedean and in urgent need of a Eureka. This paper is a step towards such an Eureka.

In many experiments IOs are driven mechanically by strings or rods, but this interferes with the natural granular structure around the IO. For example, even if the IO is pulled by an infinitely thin string, the string imposes an artificial linear void space in the moving direction, whose effect on the drag force need not be negligible. This effect, which is often ignored in the interpretation of the experimental results, must be taken into consideration. Recognising this problem is essential in studies aiming to understand the physical mechanisms behind limbless locomotion of active matter, such as `sand swimmers', i.e. animals that exhibit locomotion in sand \cite{Maetal09,Maetal11a,Shetal09,CoBo10,Maetal11c}.
Such studies must not only involve non-mechanical driving mechanisms but also model the IO as having an internal energy source. The aim in this paper is to study such active objects. 

To this end, we first construct a numerical simulation of a self-energised IO in two dimensions (2D), driving itself periodically in the horizontal direction. We use the simulation to study the effect of the horizontal driving on the IO's vertical dynamics. 
We find that whether the IO rises or sinks depends non-trivially on the amplitude and frequency of the periodic oscillation and we propose an intuitive `cavity model' to understand this behaviour. 
This minimal model is based on two competing mechanisms: accumulation of small particles under the IO, which effect rising, and fluidisation of the particle bed below the IO, which leads to sinking. Using the cavity model, we make the following quantitative predictions:
(i) the IO can rise only if the frequency is below a critical frequency, $f_c$;
(ii) the IO cannot rise if the amplitude is below a minimal value, $A_{min}$, even if the frequency is below $f_c$;
(iii) the IO cannot sink if the acceleration amplitude is below a critical value, $g_c$. 
These quantities are derived explicitly in terms of the IO and bed particle characteristics and are found to agree excellently with the numerical results.

The structure of this paper is the following. In section \ref{SimProc}, we describe the system and the simulation procedure. In section \ref{SimRes} we present the raw simulations results. In section \ref{Theory}, we present our cavity model and make quantitative predictions concerning regions in the amplitude-frequency phase diagram, where the IO cannot rise or sink. We conclude in section \ref{Conclusion} with a discussion of the results and suggested future work.

\section{\label{SimProc}Simulation procedure and parameters} 

\begin{figure}[h]
\includegraphics[width=0.5\textwidth]{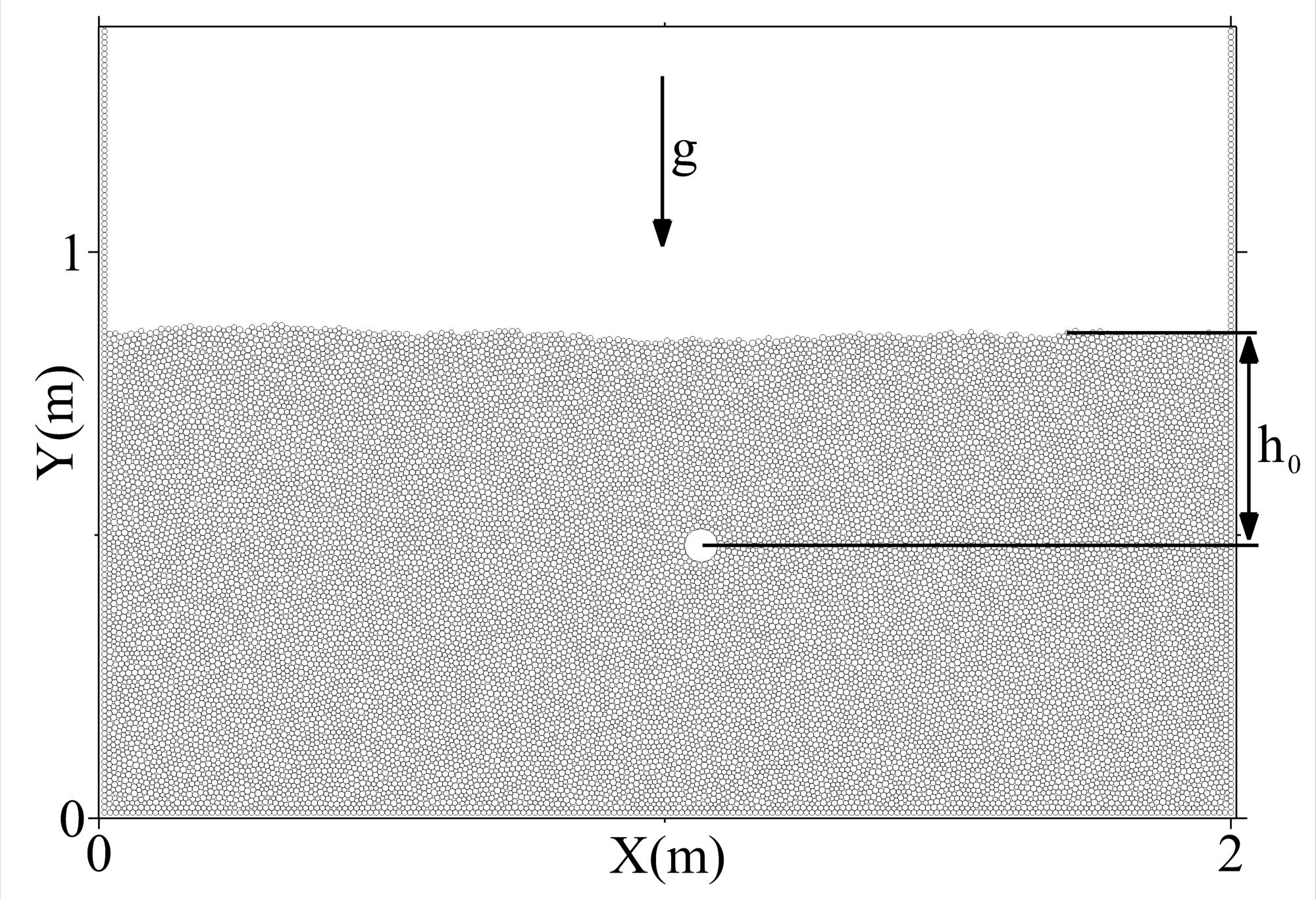}
\caption{The 2D simulation system: a large particle (white circle), of diameter $D$, is immersed at an initial depth $h_0$ and it oscillates horizontally at a velocity $v=A\sin(2\pi f t)$.}
\label{System}
\end{figure}

The simulated 2D system, sketched in figure \ref{System}, is a rectangular container of width $200l$, filled with $20,000$ discs, whose diameters are distributed uniformly between $0.8l$ and $1.2l$, such that the mean diameter is $\bar{d}=1l$. For reference, we take $l=0.01$m and measure all lengths in terms of $\bar{d}$, in the following. 
The discs are chosen to emulate glass discs of a unit thickness and of density $2500$Kg/m$^3$, giving $m\approx 1.308\times 10^{-3}$Kg.
The system also consists of one IO of diameter $D=6\bar{d}$, representing a hollow glass disc of average density $10^3$Kg/m$^3$, which gives it a mass $M=0.113$Kg. 

An initial simulation brings the discs to mechanical equilibrium in a gravitational field $g=9.81$m/sec$^2$, resulting in an assembly of average height $85\bar{d}$. The system is deemed to have reached static equilibrium when the total kinetic energy per particle gets below $10^{-15}$J.
We made sure that, after equilibration, the IO rests sufficiently away from the system's boundaries, at a depth of $h_0\approx (37\pm 0.5)\bar{d}$. This is the initial condition for all our dynamic simulations. Different initial depths are not expected to affect the dynamics, as long as the IO is sufficiently far from the top or bottom of the assembly. 

\begin{figure}[h]
\includegraphics[width=0.45\textwidth]{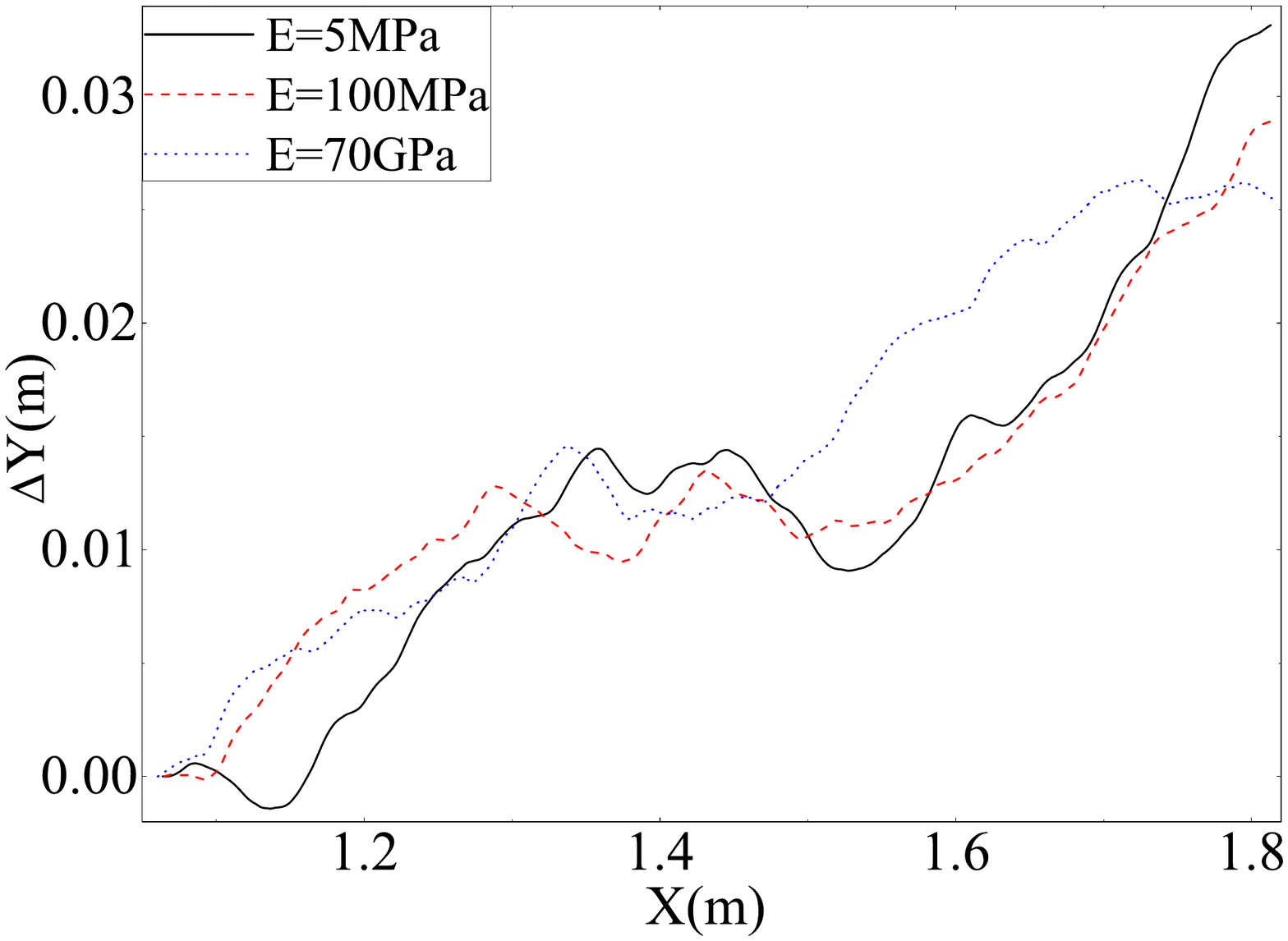}
\caption{The vertical rise of disks of mass $0.3$Kg, moving at a horizontal speed of $1$m/sec within the granular bed, for three values of Young's modulus: 
$E=5$MPa, $100$MPa and $E=70$GPa. The vertical rise of the discs at the end of the runs are similar, $(3.0\pm 0.4)10^{-2}$m, which is lower than the fluctuations during the rise process - up to $\pm 0.8$cm. Note that the fluctuations are non-systematic - different discs rise faster at different times. This establishes that, within this range, the behaviour is not sensitive to Young's modulus and we can use $E=5$MPa in our simulations.}
\label{ETest}
\end{figure}

We use the contact force model of Brilliantov et al. \cite{Beetal96}, described briefly in the following. Given two particles of diameters $d_i$ and $d_j$, moving at velocities $\vec{v}_i$ and $\vec{v}_j$ and separated by a centre-to-centre distance of $r_{ij}$, their overlap distance is defined as $\delta_{ij}=\frac{d_i+d_j}{2}-r_{ij}$. The inter-particle normal force is 

\begin{equation}
F_{ij} = 
 \begin{cases}
 0  & \delta_{ij} < 0 \\
  k_{ij}\delta_{ij}^{3/2} - a k_{ij}\delta_{ij}^{1/2} v_{n}^{rel}  & \delta_{ij} \geq 0 \ . \\
\end{cases}
\label{FNormal}
\end{equation}
Here $v_{n}^{rel}$ is their relative velocity's component parallel to $r_{ij}$, $k_{ij}=E\sqrt{2d_{eff}}/\left[3\left(1-\nu^2\right)\right]$ is often named `hardness', $E$ is Young's modulus, $\nu$ is Poisson's ratio, $d_{eff}\equiv d_i d_j /\left(d_i+d_j\right)$, $\gamma_t$ is the viscosity drag, and $\mu$ the solid friction coefficient. These parameters are chosen to mimic 3D glass discs \cite{Guetal14}: $\nu=0.45$, $\gamma_t=10$Ns/m, $\mu=0.3$ between bed discs and $\mu=0.4$ between the IO and a bed disc. 
Using the derivation in \cite{Scetal98}, of the velocity dependence of the restitution coefficient, we find that it varies by less than $0.3\%$ over the range of velocities involved in our simulations, $0.1$m/sec $\le v \le 4.0$m/sec. Consequently, the restitution coefficient can be safely taken to be constant. The parameter $a=1.5\times 10^{-6}$s is then determined by fitting the restitution coefficient in a simulation of a disc bouncing on a hard surface, to a value of 0.8, which is the typical value for glass beads found in a number of experiments \cite{Guetal14,TaMo82}.
Young's modulus was set at $E=5$MPa, which is lower than the physical value of a few tens of GPa. The reason for that is that a simulation with a more realistic value would require a time step that is $1000-10000$ times smaller than the one we could use. Indeed, this is the reason that most works in the literature use  $E=5$MPa, e.g. \cite{Guetal14,PoDi13}.
To test that this reduced value does not introduce undesired elastic effects, we ran a series of long simulations on $0.3$Kg discs of different Young's moduli, $E=5$MPa, $100$MPa and $E=70$GPa, moving horizontally at $1$m/sec within the granular bed. The vertical rise of the discs at the end of the runs were within the range $(3.0\pm 0.4)10^{-2}$m, while the fluctuations of the rise within each run were much higher, up to $\pm 0.8$cm (figure \ref{ETest}). Moreover, the fluctuations are non-systematic in that different discs rose more than the others at different times. This established that the any elastic error is negligible.
The tangential force is

\begin{equation}
F_{t} = -sign\left\{v_{n}^{rel}\right\}\times min\left(\gamma_t |v_{t}^{rel}|,\mu F_n\right)
\label{FTangential}
\end{equation}
where $\vec{v}_{t}^{rel}=\vec{v}^{rel}-\vec{v}_{n}^{rel} + d_j \vec{\phi}_i/2 - d_j \vec{\phi}_j/2$ and $\phi_k$ is the angular velocity of the $k$th particle, defined positive in the anticlockwise direction (in 2D it is always normal to $\vec{r}_{ij}$).

Starting from the equilibrated initial state, the IO is oscillated horizontally at velocity $v_x(t)=A\sin(2\pi f t)$ and the its dynamics are simulated using the Gear algorithm \cite{Ge66,Ge71} with a time step of $\delta t=10^{-6}$sec. We ran a series of DEM simulations, each for $10^7$ time steps ($=10$sec), for a range of values: $0.1$m/sec$^2  \leq A \leq 4.0$m/sec$^2$ and $1$Hz $\leq f \leq 20$Hz.
In each run we monitored the dynamics of the IO and, in particular, we measured its vertical rise at each time step. In the following, negative rise corresponds to sinking in the direction of gravity and visa versa. 

\section{\label{SimRes}Simulation results} 

\begin{figure}[!h]
\begin{minipage}[b][12cm]{0.5\linewidth}
{\includegraphics[width=3.8cm,height=3cm]{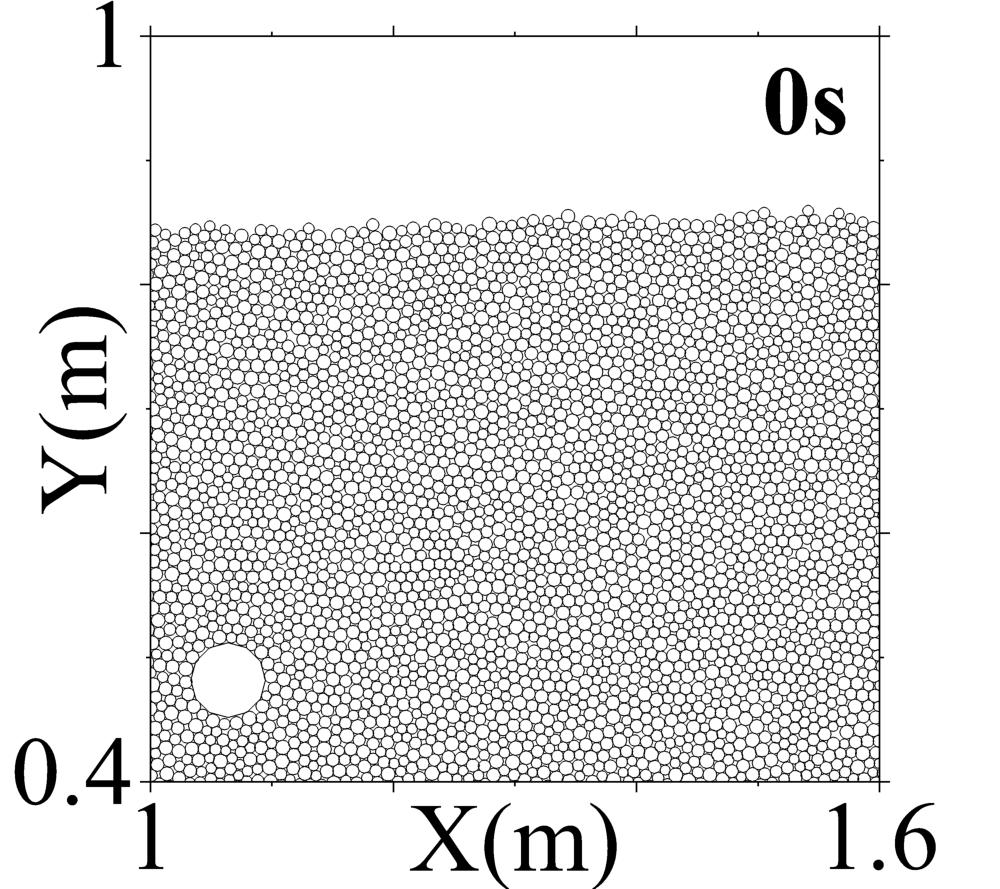}}
\vfill
{\includegraphics[width=3.8cm,height=3cm]{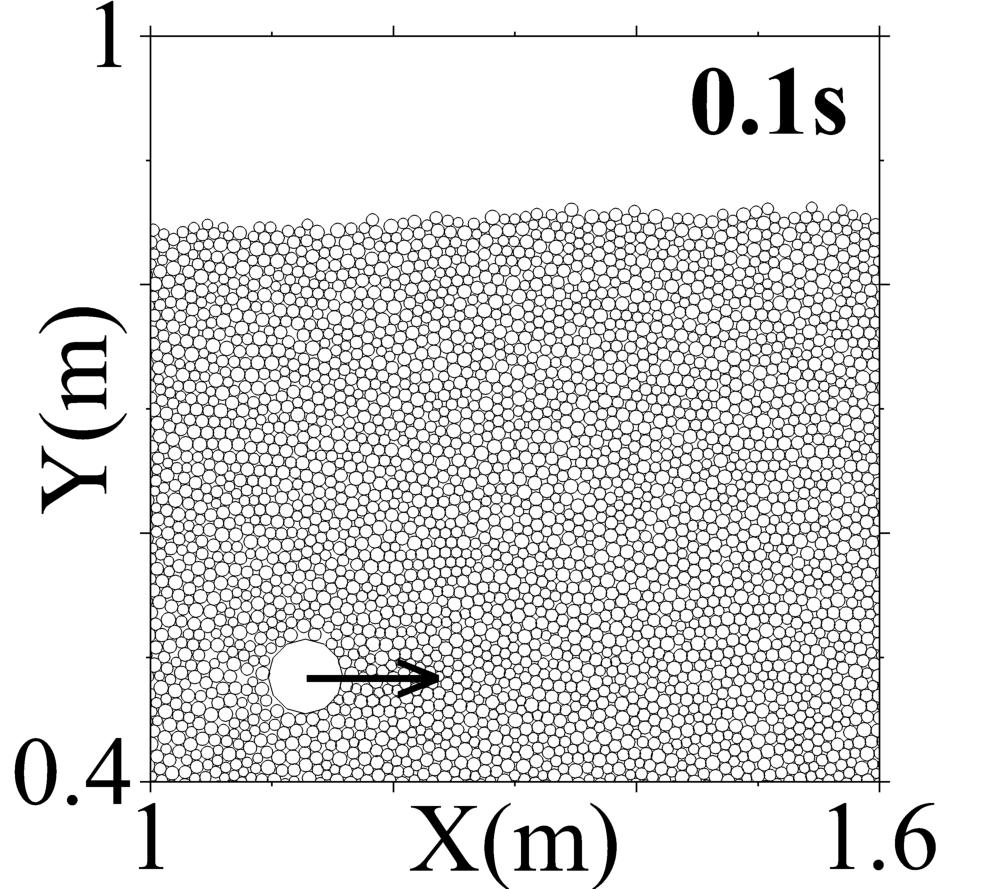}}
\vfill
{\includegraphics[width=3.8cm,height=3cm]{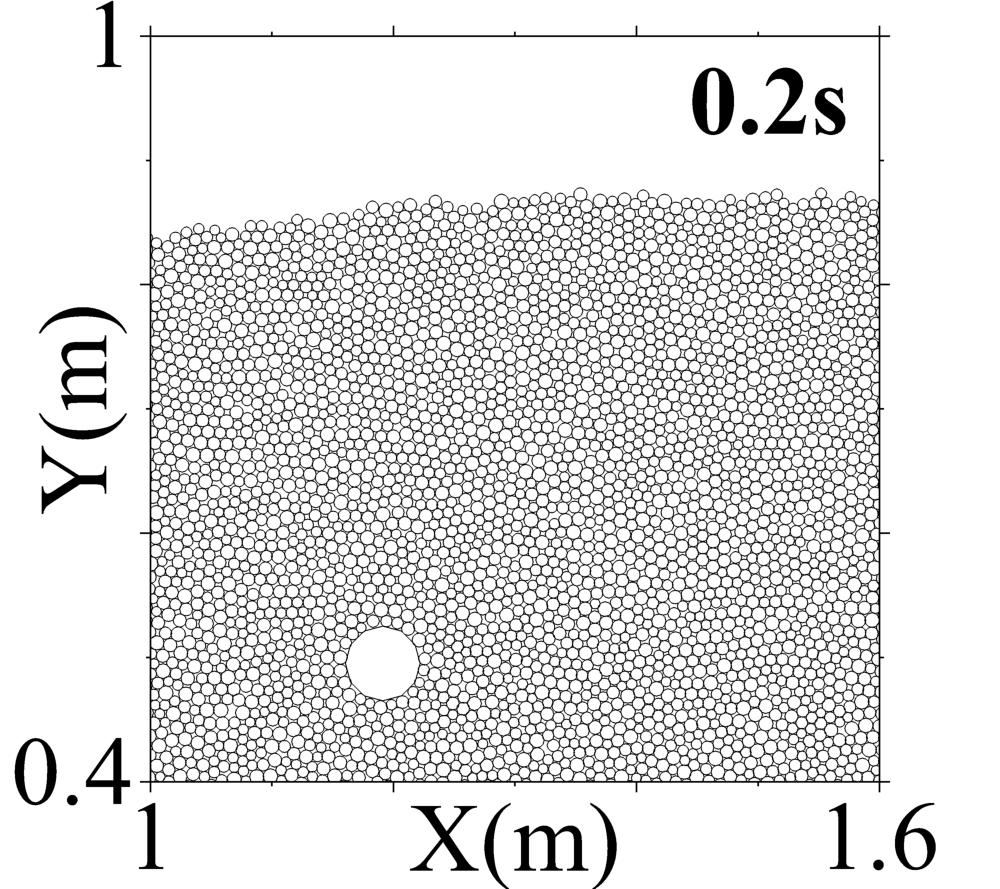}}
\vfill
{\includegraphics[width=3.8cm,height=3cm]{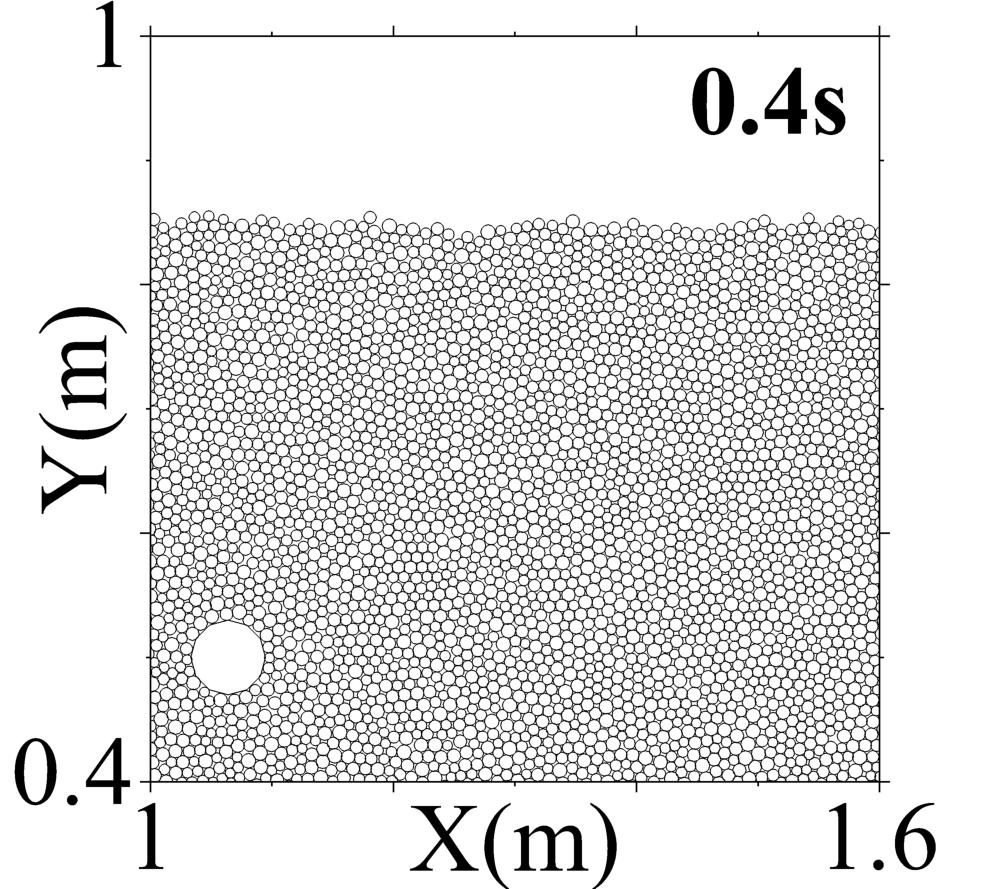}}
\end{minipage}%
\begin{minipage}[b][12cm]{0.5\linewidth}
{\includegraphics[width=3.8cm,height=3cm]{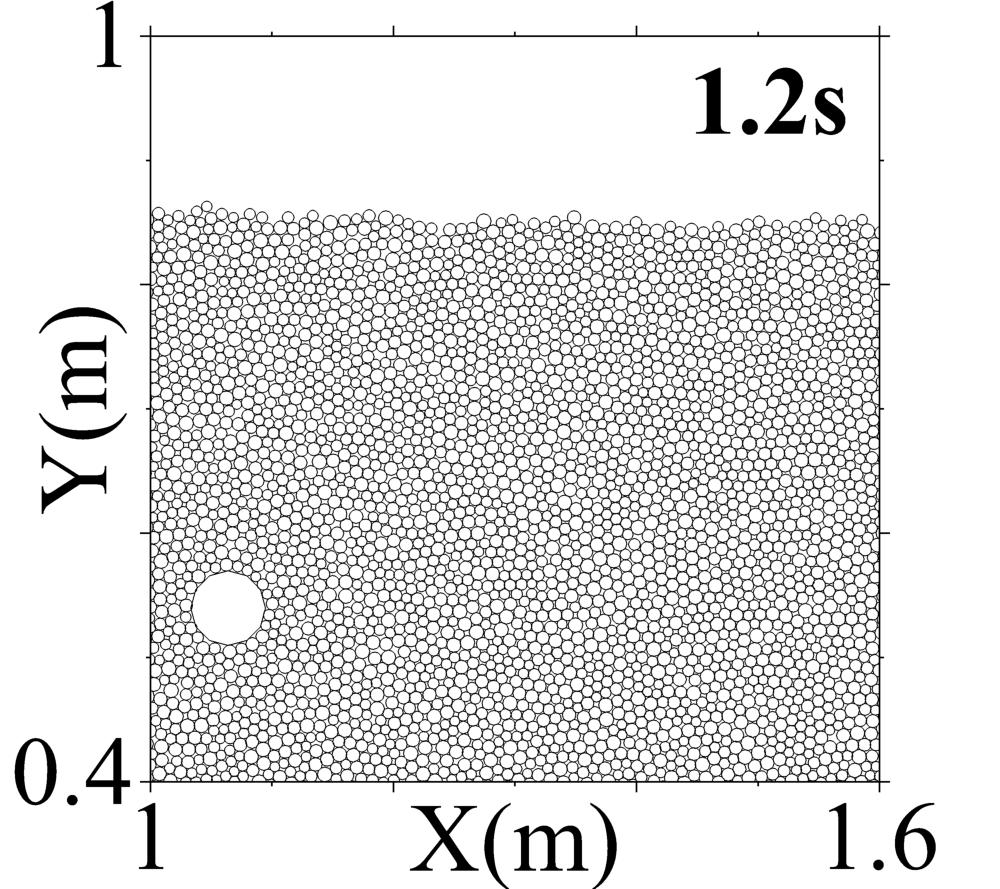}}
\vfill
{\includegraphics[width=3.8cm,height=3cm]{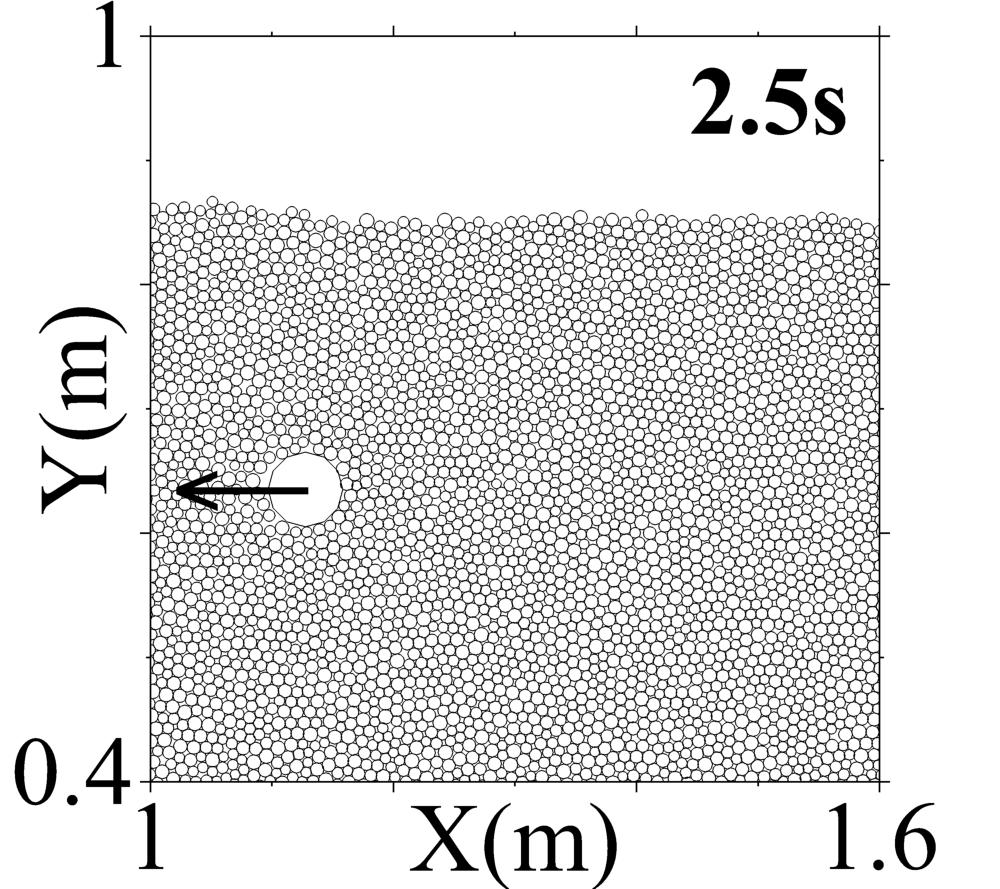}}
\vfill
{\includegraphics[width=3.8cm,height=3cm]{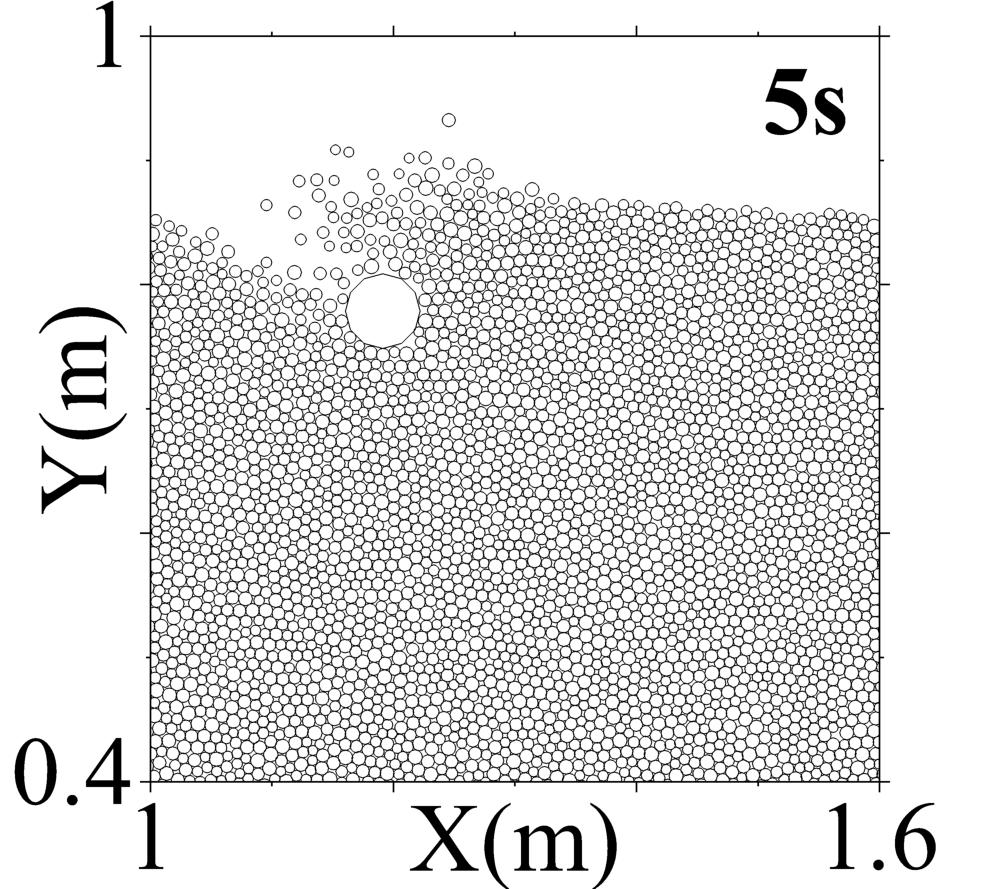}}
\vfill
{\includegraphics[width=3.8cm,height=3cm]{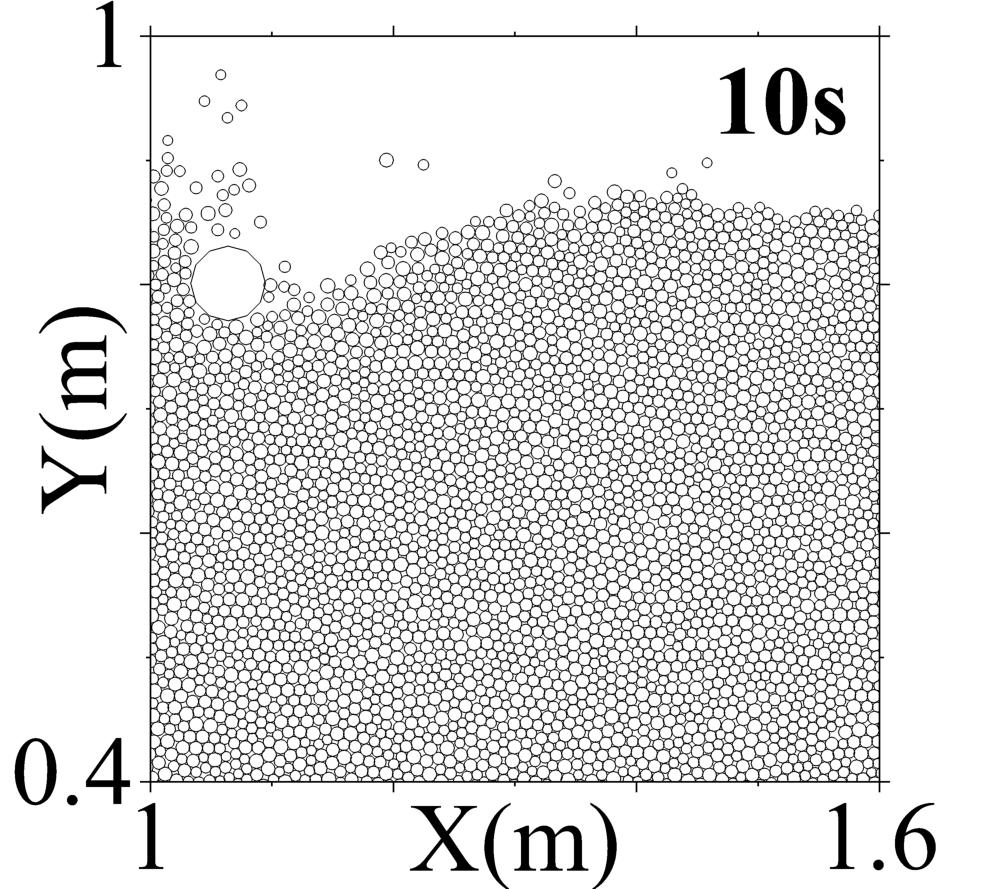}}
\end{minipage}%
\caption{An illustration of a rise process: $A = 1$m/sec and $f = 2.5$Hz. Only the relevant part of the system is shown.}
\label{fig2_RiseExamplea}
\end{figure}

\begin{figure}[!h]
\begin{minipage}[b][12cm]{0.5\linewidth}
{\includegraphics[width=3.8cm,height=3cm]{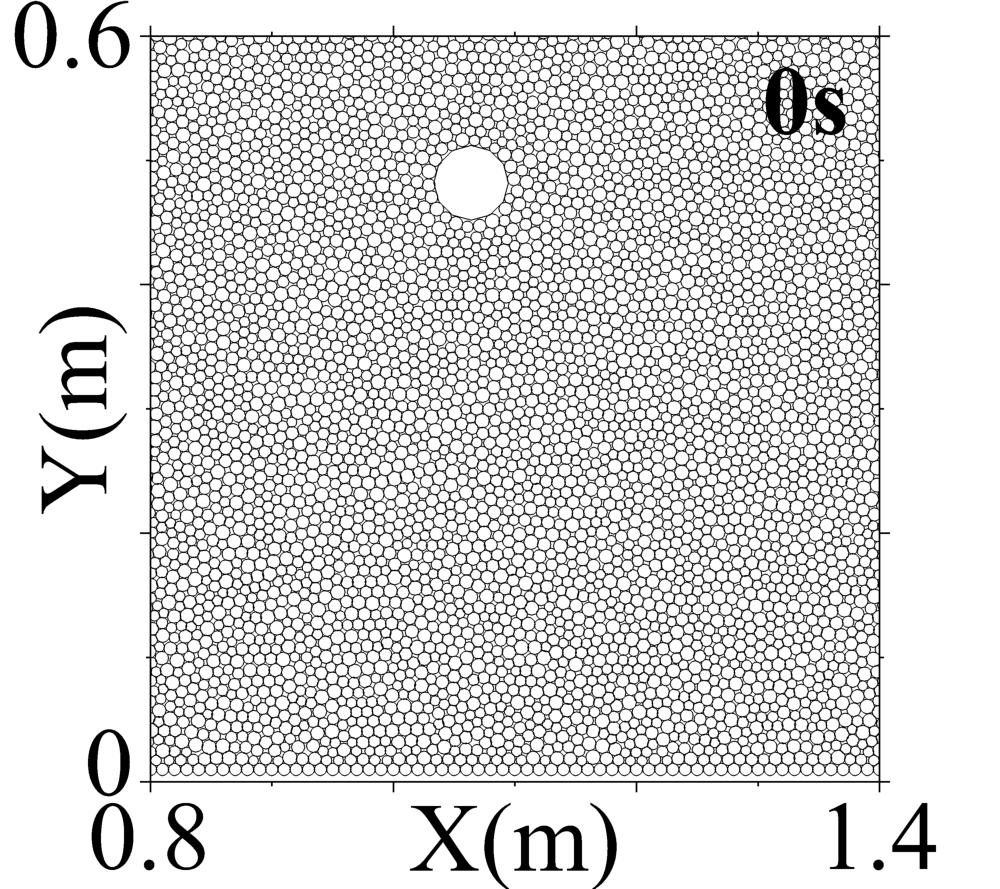}}
\vfill
{\includegraphics[width=3.8cm,height=3cm]{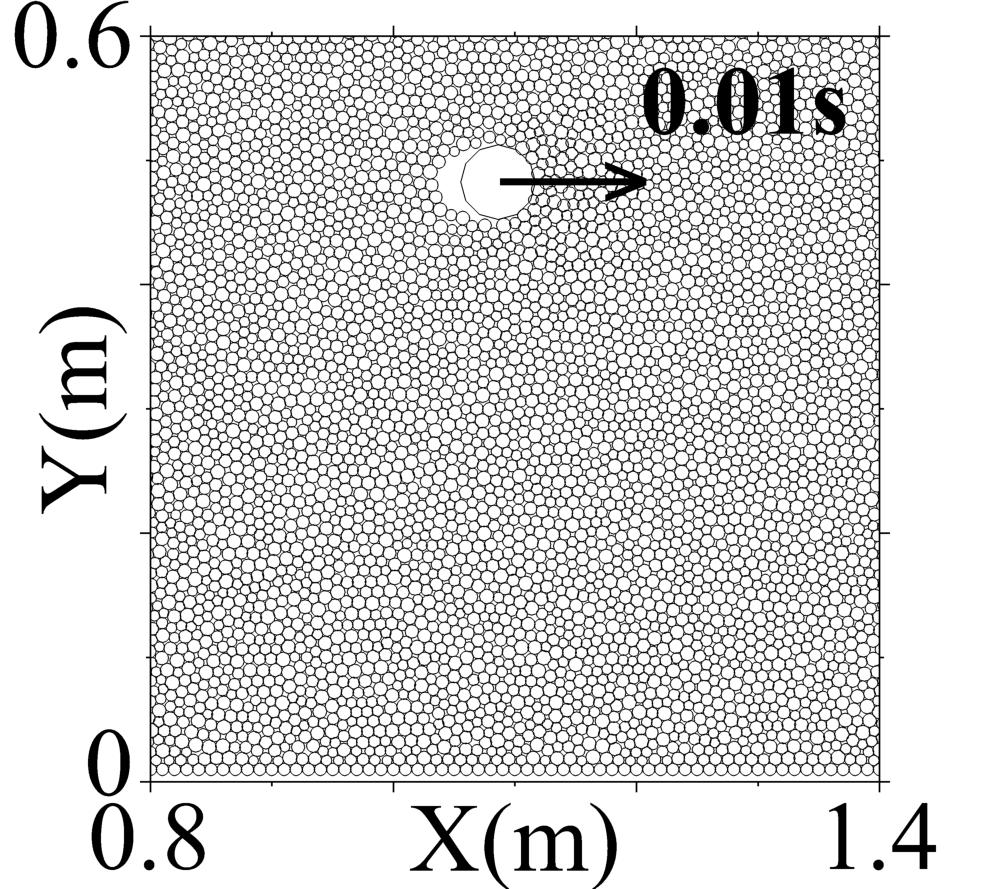}}
\vfill
{\includegraphics[width=3.8cm,height=3cm]{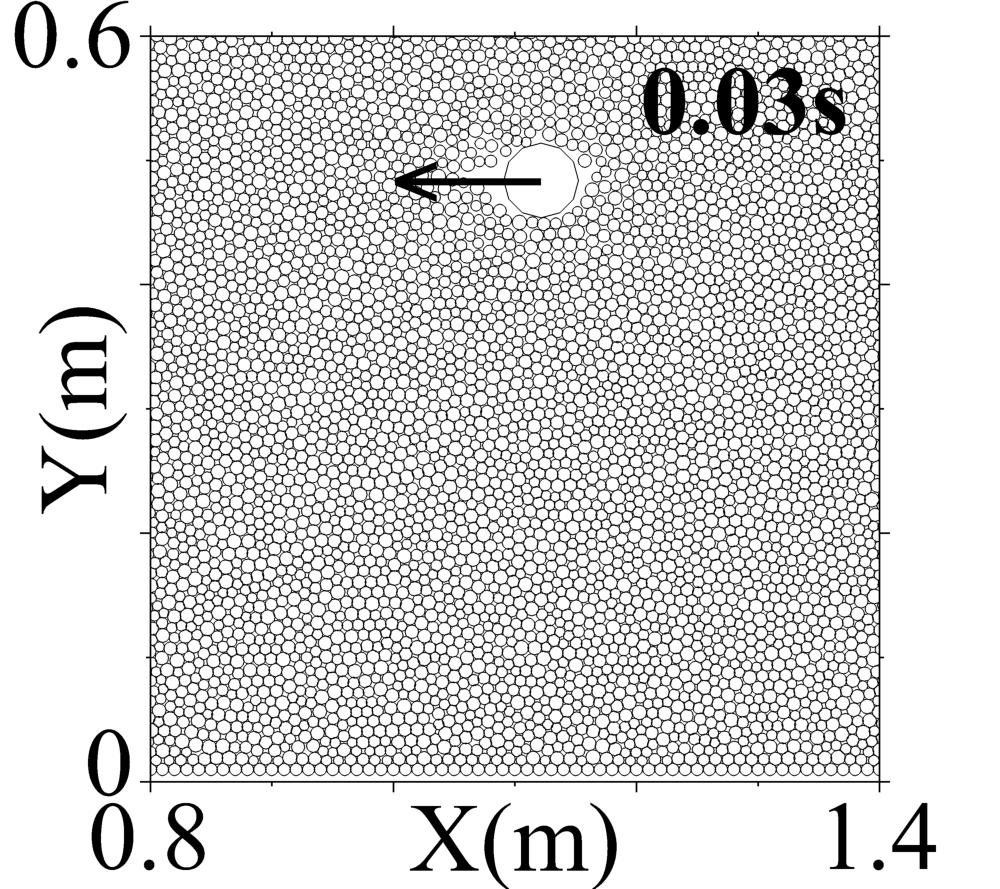}}
\vfill
{\includegraphics[width=3.8cm,height=3cm]{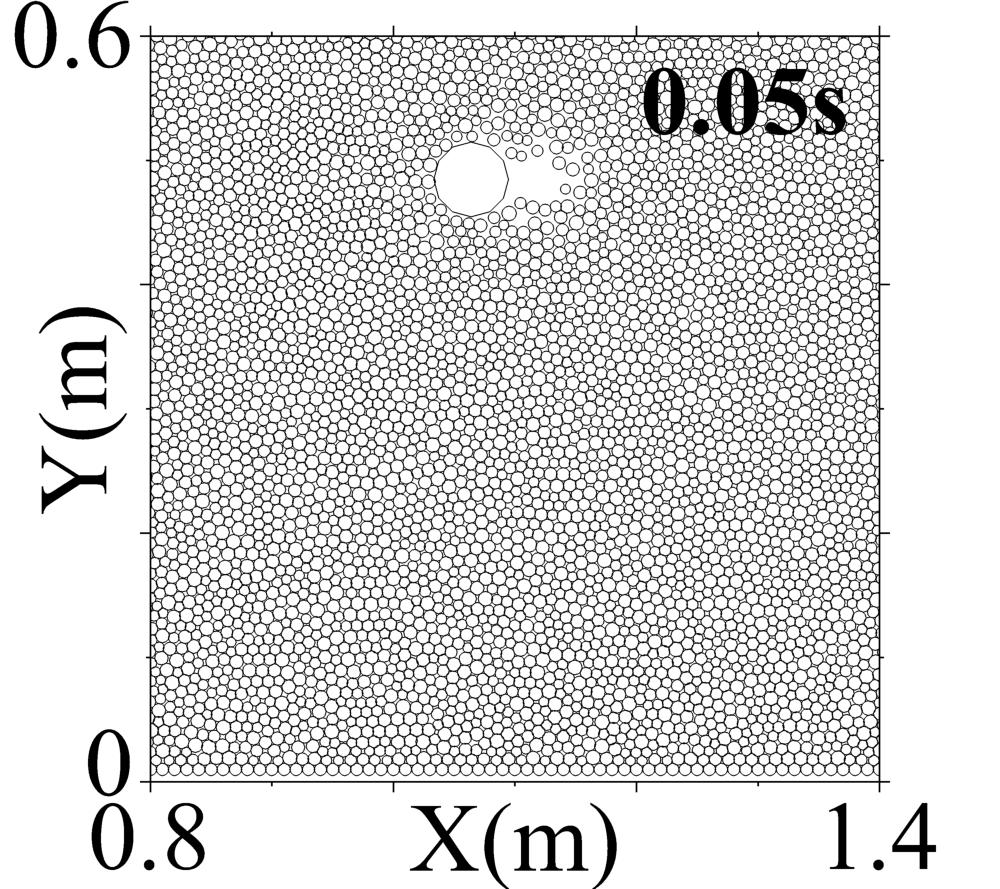}}
\end{minipage}%
\begin{minipage}[b][12cm]{0.5\linewidth}
{\includegraphics[width=3.8cm,height=3cm]{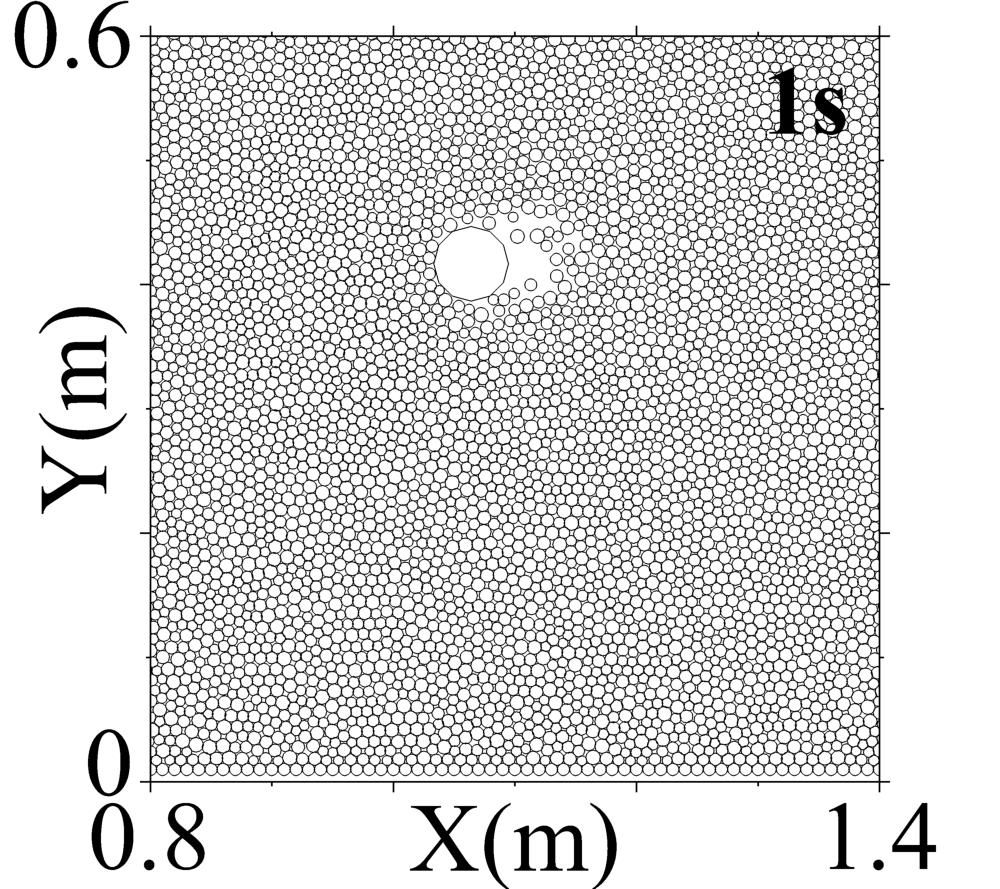}}
\vfill
{\includegraphics[width=3.8cm,height=3cm]{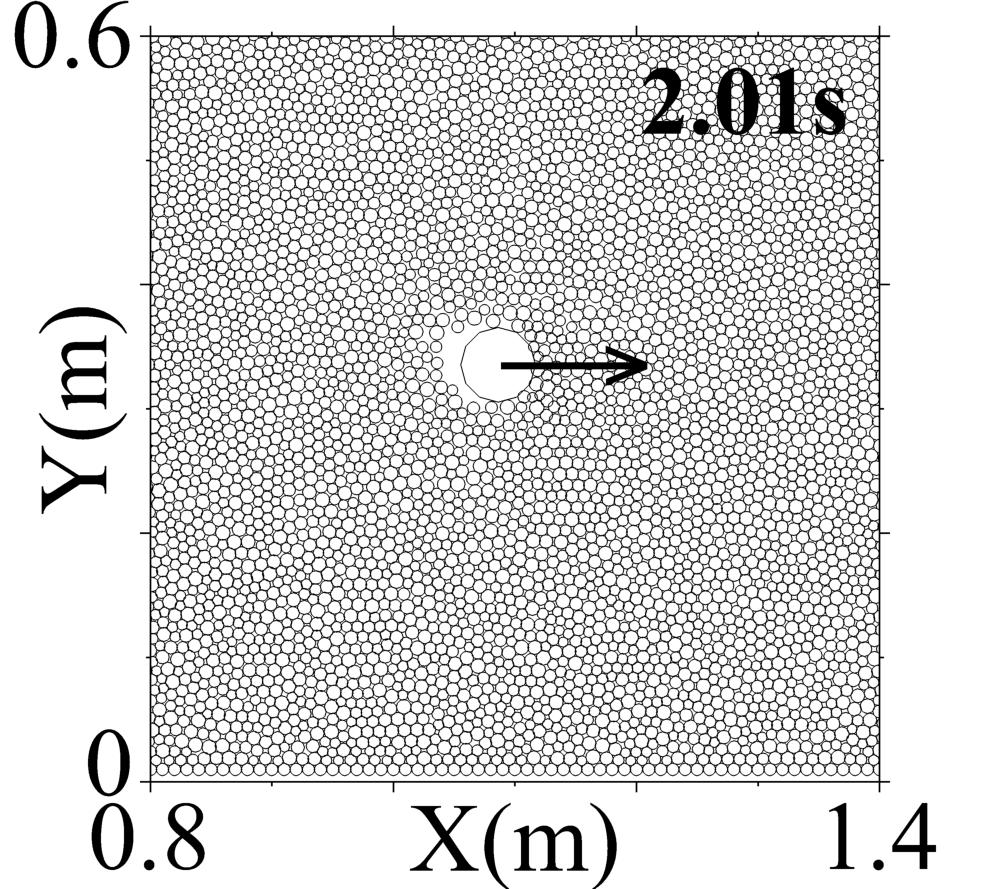}}
\vfill
{\includegraphics[width=3.8cm,height=3cm]{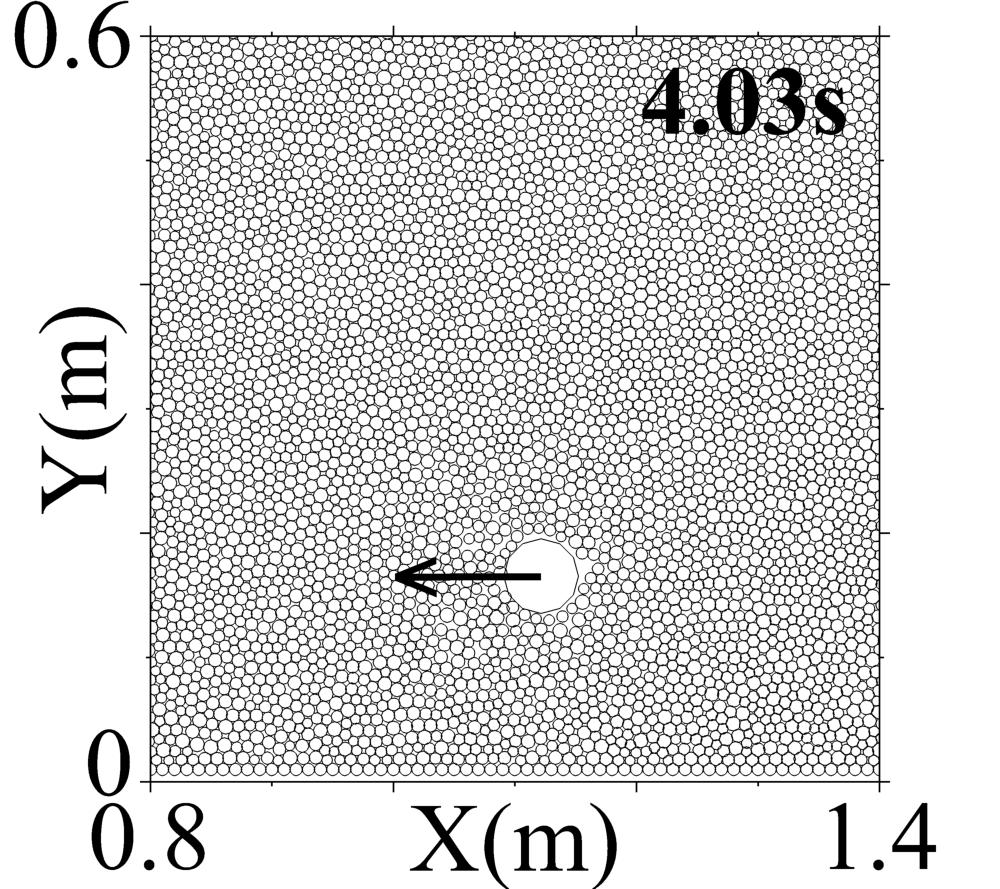}}
\vfill
{\includegraphics[width=3.8cm,height=3cm]{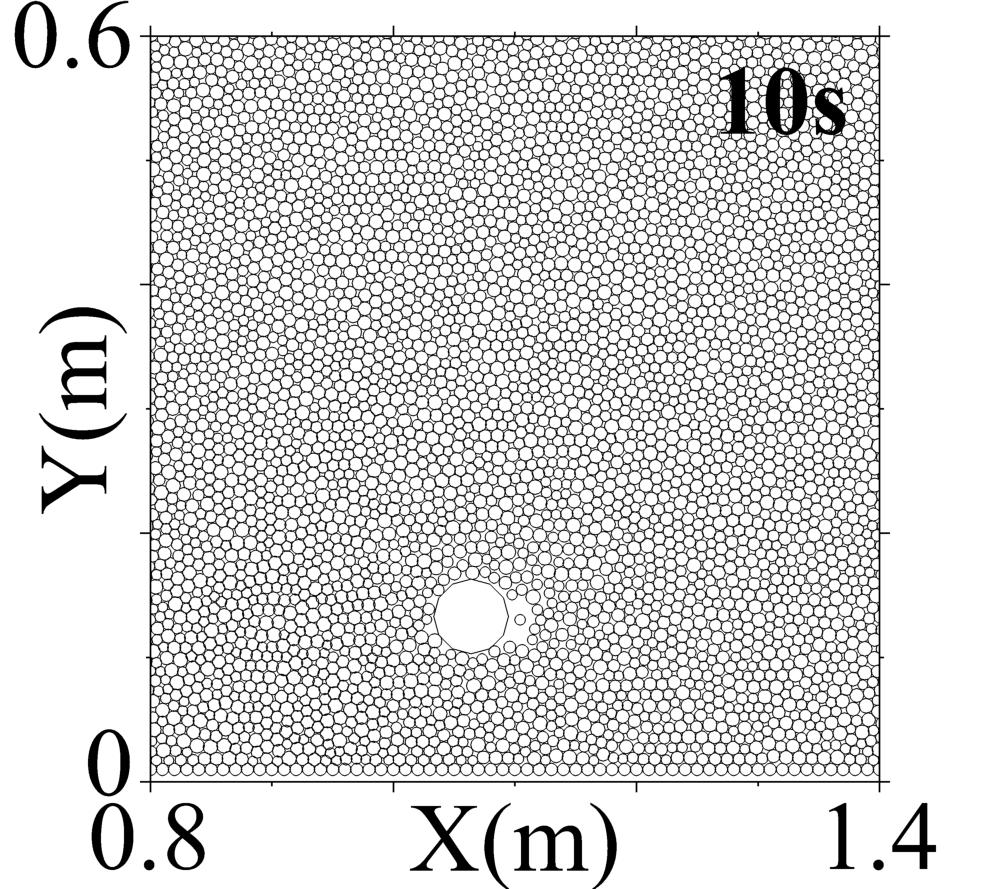}}
\end{minipage}%
\caption{An illustration of a sink process: $A = 4$m/sec and $f = 20$Hz. Only the relevant part of the system is shown.}
\label{fig3_SinkExampleb}
\end{figure}

\begin{figure}[!h]
{\includegraphics[width=0.5\textwidth]{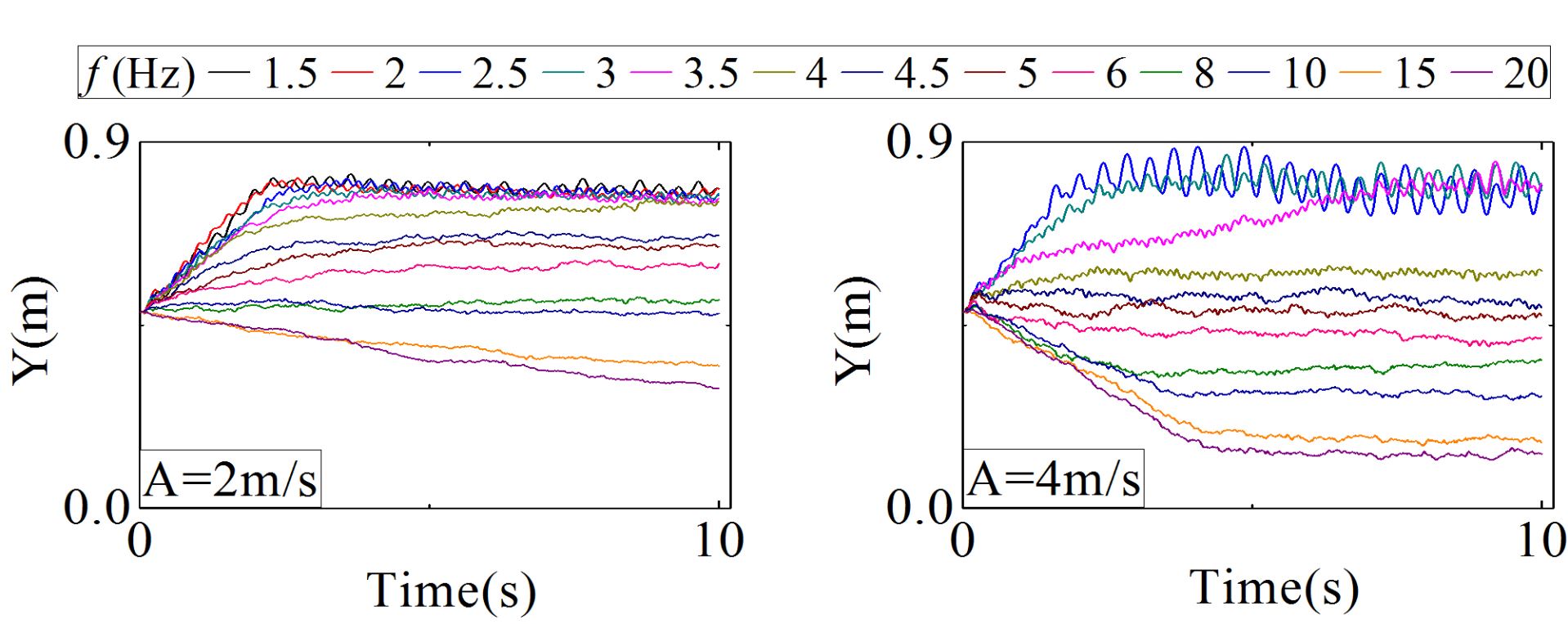}}
\caption{Left: The rise, as a function of time, for amplitude $2$m/sec and all frequencies. Right: The same for amplitude $4$m/sec. The settling level at the top of the figures corresponds to the IO reaching the top surface.}
\label{fig4_RiseTimeSeries}
\end{figure}

\begin{figure}[!h]
{\includegraphics[width=0.5\textwidth]{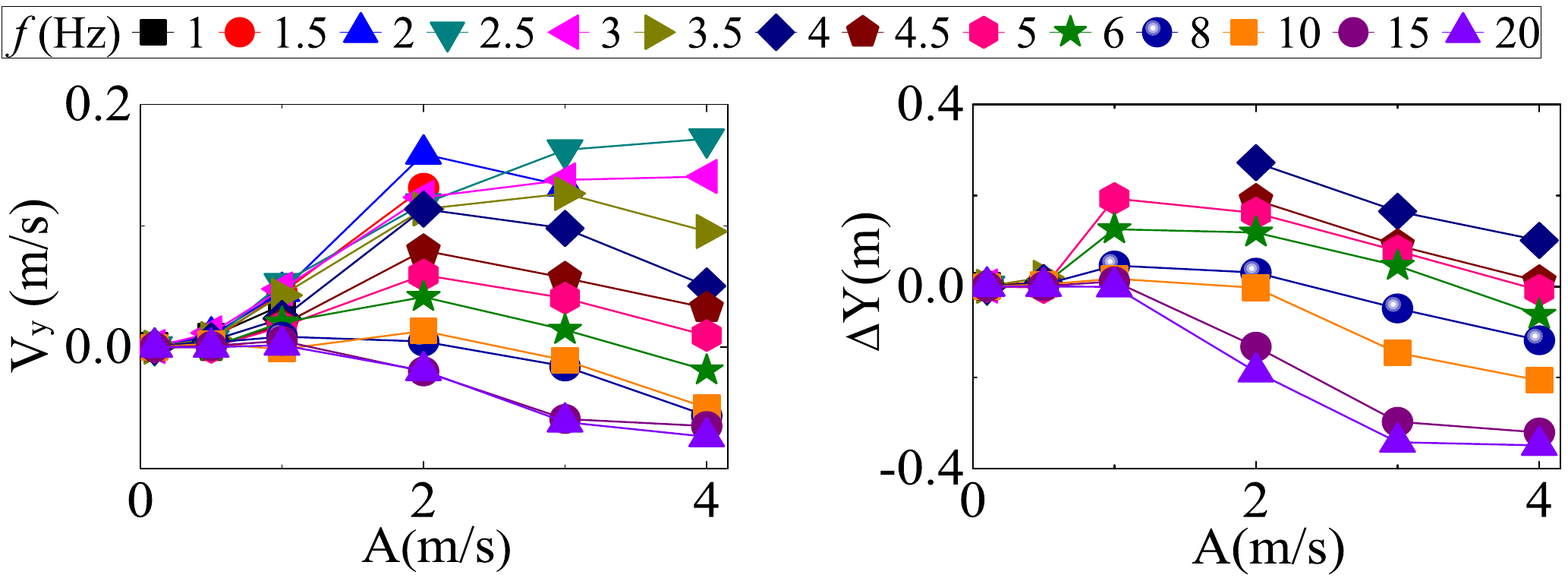}}
{\includegraphics[width=0.5\textwidth]{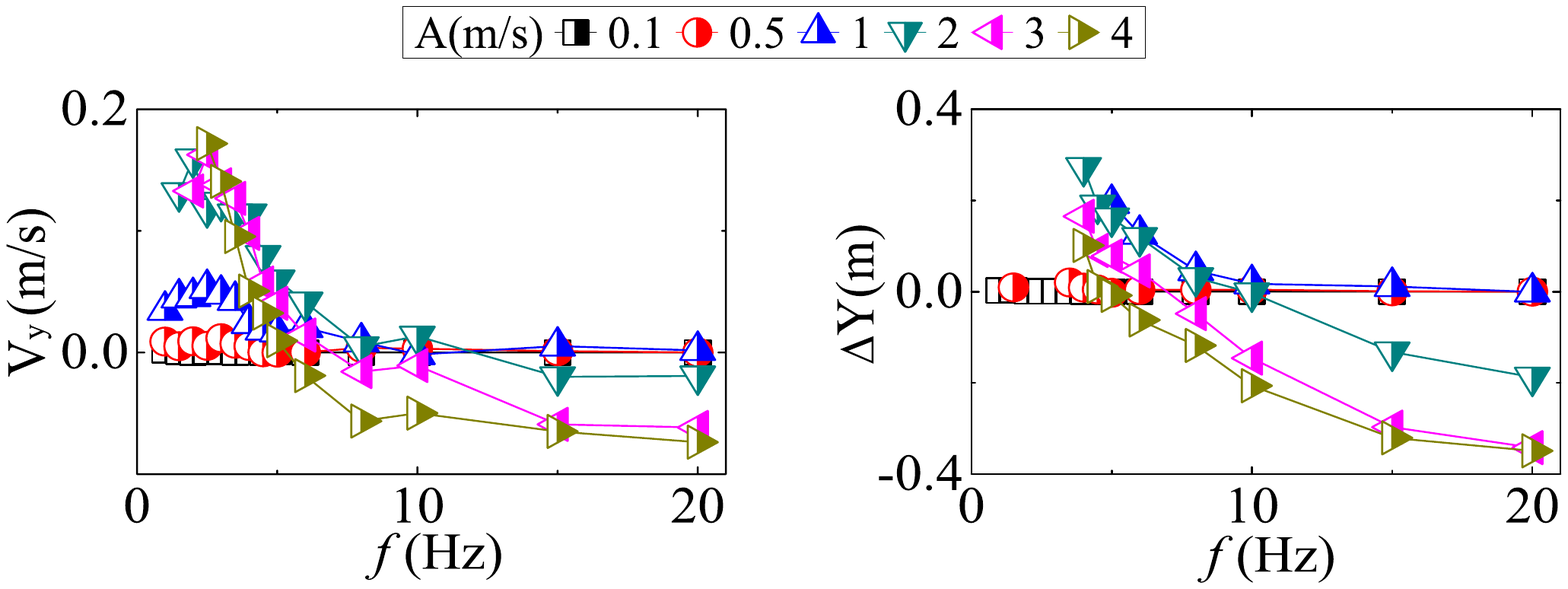}}
\caption{Top left: the initial rise rate, plotted as a function of velocity amplitude, for all frequencies. Bottom left: the same, but as a function of frequency, for all velocity amplitudes. 
Top right: the settling depth, as a function of velocity amplitude, for all frequencies, excluding the cases when the IO either reached the top surface or did not settle before the simulation ended. Bottom right: the same processes as top right, but as a function of frequency, for all velocity amplitudes.}
\label{fig5_RiseRateSettle}
\end{figure}

For every pair of values, $A$ and $f$, we measured the height of the IO, as a function of time. We observed the following features: the rise rate, $v_y=d(\Delta Y)/dt$, can be either positive or negative, it depends sensitively on the oscillation amplitude and frequency, it is not constant, and the IO eventually settles at a specific depth level, $\Delta Y_\infty$, which we will denote simply $\Delta Y$, for brevity, except where confusion may arise. These phenomena are illustrated in figures \ref{fig2_RiseExamplea} and \ref{fig3_SinkExampleb}, where examples of two opposite rise rates are shown. 
In figure \ref{fig4_RiseTimeSeries}, we show two examples of rise processes for amplitudes $2$m/sec and $4$m/sec and all frequencies. 
The initial rise rates, i.e. before any settling time, for all the cases, is shown in figure \ref{fig5_RiseRateSettle} (top left and bottom left).

The value of the final settling depth was also found to depend sensitively on the oscillation parameters, $A$ and $f$. This is illustrated in figures \ref{fig4_RiseTimeSeries} and \ref{fig5_RiseRateSettle} (top and bottom right), which give the final depths for all frequencies as a function of $A$ and for all amplitudes as a function of $f$. 
Unfortunately, we could not use all the data for analysing the final settling. Some of the rise rates were too high, such that the IO reached the top surface of the medium during the simulation. The fastest this happened was after $1.5\times 10^6$ time steps, corresponding to $t=1.5$sec. 
Additionally, in some cases, the IO rose or sank too slowly to settle into a final depth before the end of the simulation. These were excluded from the relevant analyses of the final depth, to be discussed below.

In figures \ref{fig5_RiseRateSettle}, we show the initial rise rates and the final depths, which the IO settles into, as functions of the frequency, for all amplitudes. 
Again, the latter excludes the cases when the IO either reached the surface or did not settle before the simulation ended. 
The data reveal several systematic features: 
(i) the IO never sinks for oscillations with amplitudes $A\leq 1$m/sec, regardless of frequency; 
(ii) the IO always rises for oscillations with frequencies $f\leq 10$Hz, except when the amplitude is below a minimal value of $1$m/sec;
(iii) when the IO sinks, for $A>1$m/sec and $f>10$Hz, this is always associated with formation of a large cavity in the wake of the IO's horizontal motion, already within the first oscillation period. 

\begin{figure}[!h]
\includegraphics[width=0.5\textwidth]{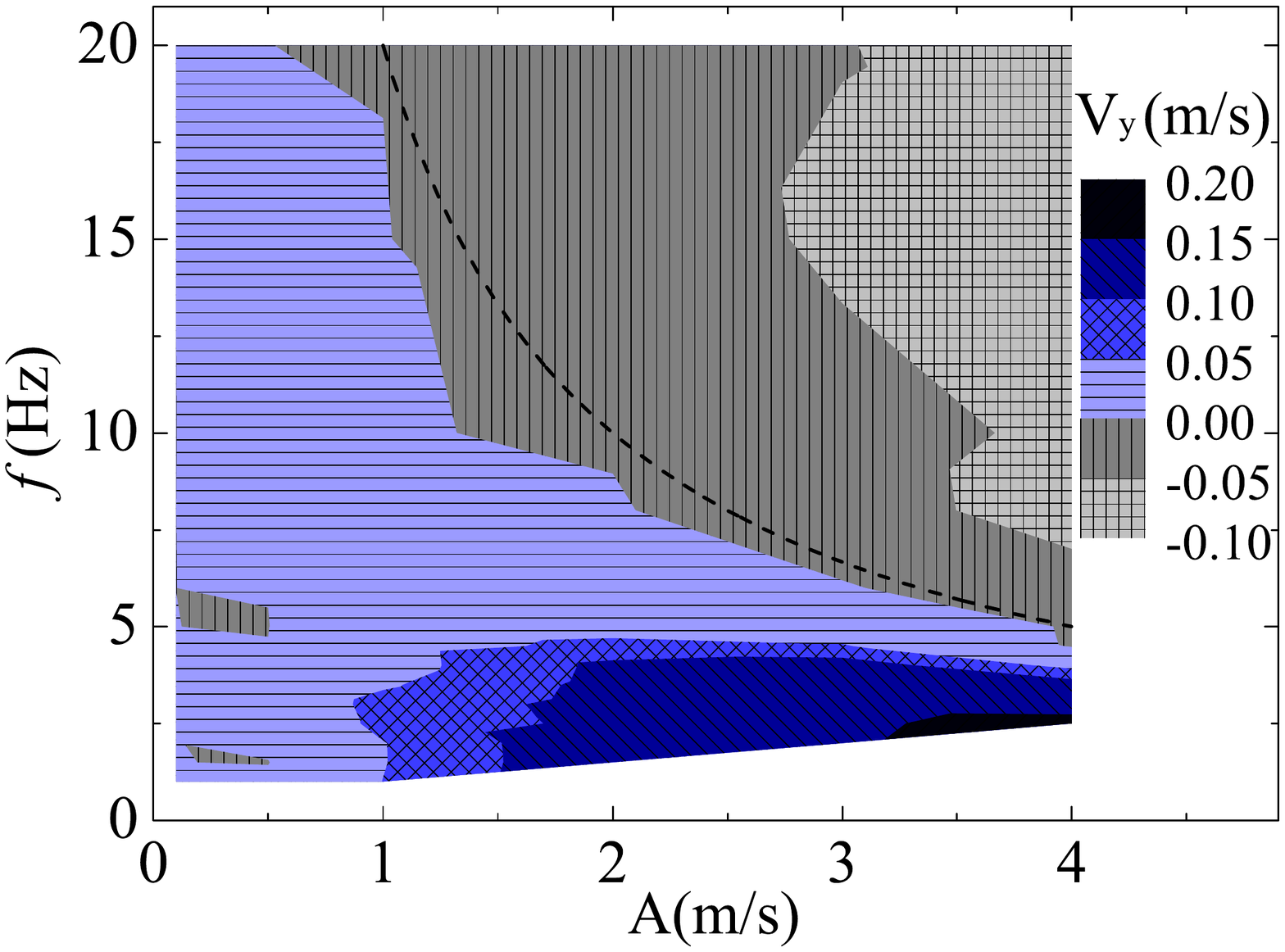}
\caption{The phase diagram of the IO rise rate for all frequencies, $f$, and velocity amplitudes, $A$. The dashed line is our prediction (see the theoretical section) of the phase boundary $g_c=(120\pm 5)$m/sec$^2$, below which no sinking is possible.}
\label{fig6_PhaseDiagramSettlingDepth}
\end{figure}

From figure \ref{fig4_RiseTimeSeries}, we see that if the IO started rising initially then it continues rising and visa versa. Therefore, we plot in figure \ref{fig6_PhaseDiagramSettlingDepth} a phase diagram of the initial rise, in the phase space of $A$ and $f$. This phase diagram follows closely the phase diagram of the final depth. 
As can be observed, the behaviour is quite rich. Nevertheless, it can be understood quantitatively, as we show in the next section.

\section{\label{Theory}Theory - the cavity model} 

We propose that two main competing physical mechanisms govern the vertical rise. One is the climb of the IO on particles that accumulate at the bottom of the void, left in its wake. As it moves through one stroke of the oscillation, the IO leaves behind a cavity, which may fill with particles before it returns at the next stroke. The IO then has to `climb' on top of these particles.
The second mechanism involves fluidisation of the granular bed, which supports the IO, causing it to sink through it. 
Thus, the key to understanding the rich behaviour of the IO is in capturing the effects of these mechanisms and how they depend on the parameters of the bed, the characteristics of the IO and the periodic oscillation frequency and amplitude. 

\begin{figure}[!h]
\includegraphics[width=0.5\textwidth]{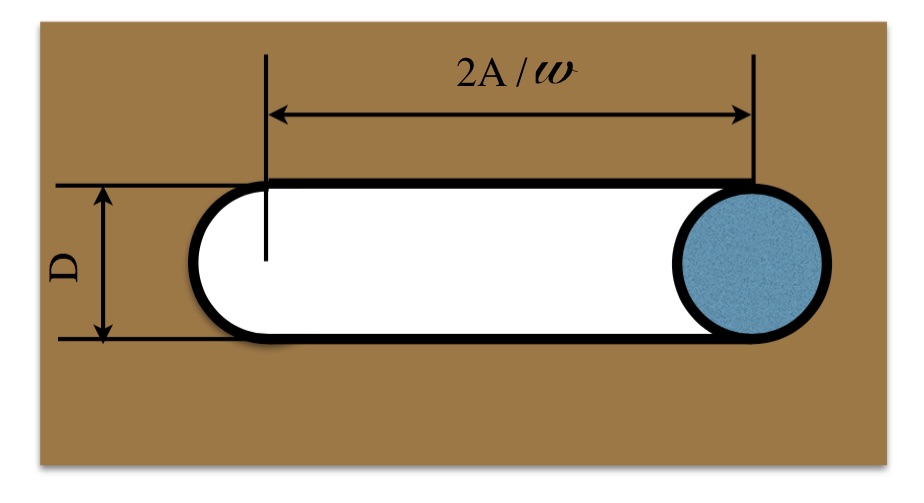}
\caption{The cavity model: as the IO moves to the right, it may leave behind a cavity, which may be filled by the surrounding particles, depending on $A$ and $f$.}
\label{fig_Cavity}
\end{figure}

Starting with the cavity picture, we idealise it in 2D as shown in figure \ref{fig_Cavity}. During the half period of one stroke, $T/2$, the rise of the IO depends on the thickness of the layer of bed particles, which manage to accumulate on the base of the cavity before it returns. 
To estimate this number, we note that the number of particle forming one layer of the top of the cavity is approximately $(2A/\omega)/\bar{d}$. The time it takes one such a layer to free-fall to the bottom of the cavity is about $\sqrt{2D/g}$, regardless of $\bar{d}$. Therefore, during one stroke of duration $T/2$, the total number of particles that fall into the cavity is, up to a numerical factor of order 1, 

\begin{equation}
n_{top} = \frac{2A/\omega}{\bar{d}}\frac{T/2}{\sqrt{2D/g}} \ ,
\label{nTop}
\end{equation}
Meanwhile, the particles forming the left wall of the cavity also avalanche into it. For simplicity, we assume that this contributes a similar number of particles, $n_{av} + n_{top} \approx \alpha n_{top}$, with $\alpha$ a numerical factor of about $2$.
The thickness of the layer forming on the cavity base is then 

\begin{equation}
W=\frac{\left(n_{av} + n_{top}\right)\bar{d}}{2A/\omega}\bar{d} =  \frac{\alpha}{2}\sqrt{\frac{g}{2D}} T \bar{d} \ .
\label{LayerThickness}
\end{equation}
During this period, the IO rises by `wedging' on top of this layer \cite{Dietal11}, gaining a height of $W\tan{\theta}$. ${\theta}$ is the wedge angle, which we take to be about $45\degree$ here \cite{Dietal11}, and $\bar{d}\tan{\theta}$ is the rise over a one particle thick layer.
With two strokes per period, the rise rate is then $v_{rise}\approx \alpha\sqrt{g/2D} \bar{d}\tan{\theta}$. 

Now, the IO would rise only if, during any one stroke, there accumulates a layer that is at least one particle thick. This translates into the condition $v_{rise}T/2\geq \bar{d}\tan{\theta}$, which can be inverted to give 

\begin{equation}
f \leq f_c = \frac{\alpha}{2}\sqrt{\frac{g}{2D}}  \ .
\label{YRise0}
\end{equation}
 Taking $\alpha\approx 2$ and substituting for the IO size, which we use in our simulations, we find $f_c=9.04 Hz$. This is in excellent agreement with the observation in figure \ref{fig5_RiseRateSettle} that the IO did not float up, for any choice of parameters, when the frequency was above $\sim 10 Hz$, regardless of the oscillation amplitude. 
The phase diagram, shown in figure \ref{fig6_PhaseDiagramSettlingDepth}, also shows clearly that no rise takes place for frequencies above $f_c$. 
Note that this critical frequency depends only on gravity and the characteristics of the IO, its size $D$ and its shape-dependent wedging angle $\theta$, and it is independent of the properties of the bed particles.
In particular, the independence of $\bar{d}$ is because it requires at least one layer of bed particles to fall into the cavity for the IO to rise at all and the thickness of this layer is immaterial for this condition. Since the time it takes such layer to form depends only on the height of the cavity, $D$, and on gravity then $f_c$, which is determined by this time, also does not depend on $\bar{d}$. Note that the rise rate, $v_{rise}$, does depend on $\bar{d}$.
Relation (\ref{YRise0}) provides us with a dimensionless number, $\Gamma\equiv \sqrt{2D/g}f/2\alpha$, which is essentially the ratio between the time it takes to raise the cavity base by one particle thick layer and half the oscillation period.

The above assumes that particles do fall into the cavity, but this is not guaranteed. For the bed particles to fall at all and get below the IO, a cavity must form in the first place. This requires that, at some stage of the stroke, the IO velocity be higher than the velocity of the falling bed particle, such that they have time to reach the cavity base, as long as the frequency is lower than $f_c$. The maximal possible IO velocity is $A$ and this condition translates to 

\begin{equation}
A\geq A_{min}\approx D/\sqrt{2D/g}=\sqrt{gD/2} \ .
\label{Amin} 
\end{equation}
In other words, the IO cannot rise for velocity amplitudes below $A_{min}$ even if the above condition on the frequency is met. 
For our IO, we calculate $A_{min}\approx 0.54m/sec$. This prediction is also in excellent agreement with the simulation results: for frequencies below $f_c$, the IO rises when $A=1m/sec$ and there is no rise (nor is there any sink) for $A=0.1m/sec$ and $0.5m/sec$. This can also be clearly seen on the left side of the phase diagram shown in figure \ref{fig6_PhaseDiagramSettlingDepth}. Interestingly, this bound is also independent of the bed particles properties. 
Relation (\ref{Amin}) provides us with another dimensionless number, $\Omega\equiv \sqrt{2/gD}A$.
Note that $\Omega$ has a straightforward interpretation: it is the ratio between the time it takes a bed particle to fall the height of the cavity, $\sqrt{2D/g}$ and the time it takes the IO to move its own diameter, $D/A$.

It should be noted that the amplitude and frequency required for rising are quite low, corresponding to IO motions that are slow relative to the time that it takes the bed particles to fall from the `top' and `side wall' into the space vacated by it. Consequently, it is difficult to observe a substantial cavity volume in the simulations in this range of parameters. Nevertheless, within the range of values identified above, the falling particles manage to reach below the depth of the centre of the IO, thus causing it to climb over them on the return stroke. 

Turning to the physics of the sinking, we hypothesise that the IO sinks because the particles supporting it are fluidised by the motion. To fluidise the particles at the cavity base, the IO must impact them with sufficient energy. The IO's kinetic power at time $t$ is the product of the force it carries, $MA\omega\cos{(\omega t)}$, and its velocity $A\sin(\omega t)$. Integrating this power over two quarters of a period (keeping it positive, of course), gives exactly $M A^2$, which is the kinetic energy that it imparts on the medium during one stroke. Only a fraction $\epsilon$ of this energy is spent on fluidising the cavity base particles and their neighbours; the rest goes to push other particles out of the way and to overcome dissipation. We postulate initiation of fluidisation as the state in which individual cavity base particles move at least a distance comparable to their own size, $\bar{d}$. The IO moves a distance of $2A/\omega$ during one stroke, affecting $2A/(\omega\bar{d})$ base particles. Each such particle requires, on average, a force $\phi$ to move against friction with its neighbours. The total work required to move these particles a distance $\bar{d}$ is then $U = 2A \phi/\omega$. 
The condition for fluidisation is then $\epsilon MA^2 \geq U$, which can be written as
 
\begin{equation}
A\omega \geq \frac{2\phi }{\epsilon M}  \ .
\label{SinkCondition}
\end{equation}
The right hand side of this relation involves only material properties, whilst the left hand side involves only our control parameters, 
This relation gives us a third dimensionless number, $\Lambda\equiv \left(2MA\omega\right)/\left(\epsilon\phi\right)$, which is essentially the ratio of the force generated by the IO's motion and the force required to mobilise a bed particle a distance comparable to its own size. 
An examination of the simulation results shows that, indeed, for all amplitudes and frequencies, sinking took place only when the value of $A\omega$ exceeded a critical value $g_c=(120\pm 5)m/sec^2$.

This result also explains why the IO did not sink for amplitudes of $1m/sec$ and lower: the highest frequency we used was probably too low to fluidise the bed for such low amplitudes. This interpretation is also borne out by direct observations in the numerical simulations that sinking was always accompanied by large motion of the cavity base particles, lending support to this argument.

\begin{figure}[!h]
\includegraphics[width=0.5\textwidth]{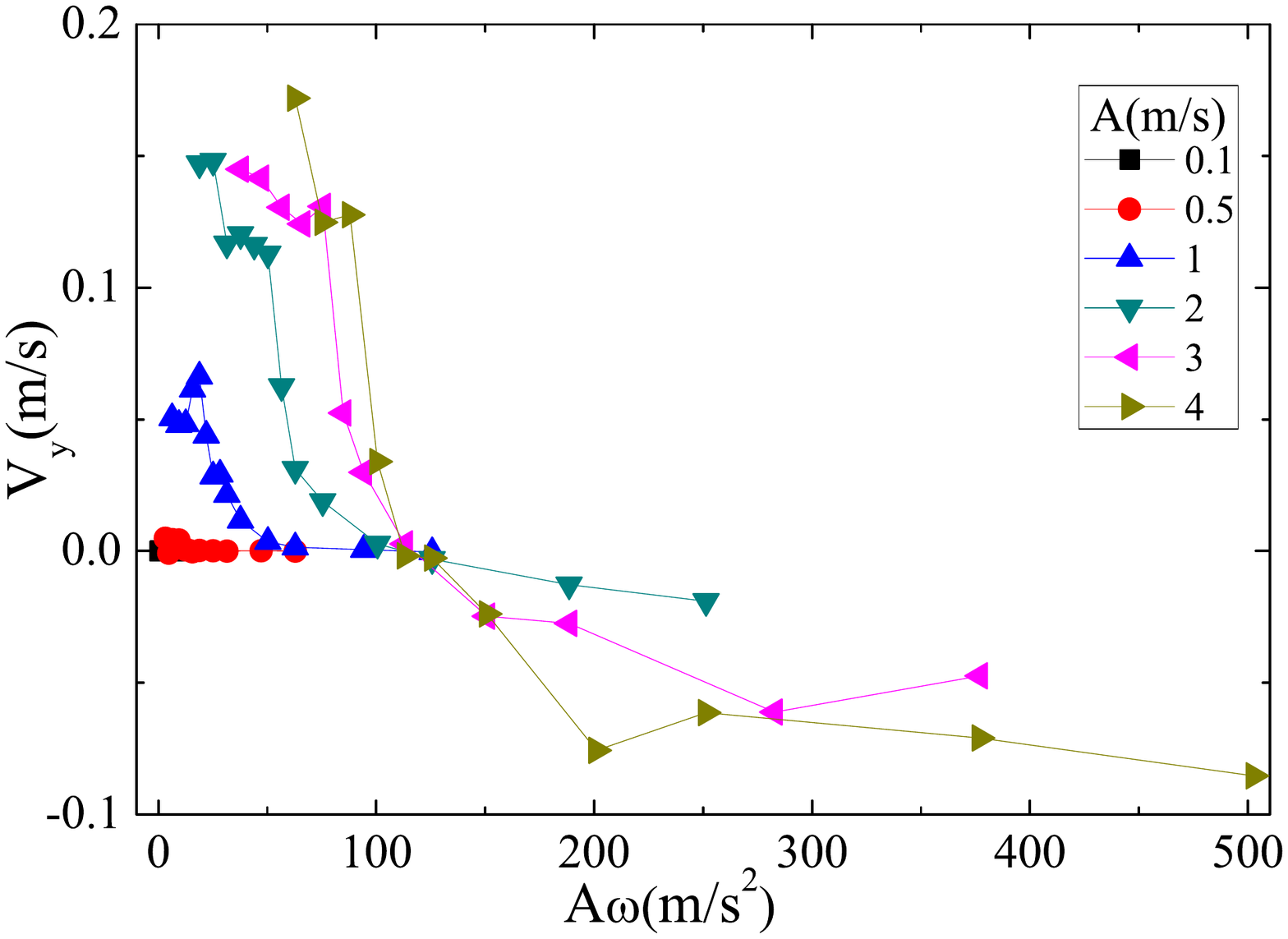}
\caption{The velocity as a function of $A \omega$ for all cases. Note that all plots converge at the point $(\Delta Y, A\omega) = (0,g_c)$, with $g_c=(120\pm 5)m/sec^2$.}
\label{fig7_VelocityVsAw}
\end{figure}

To substantiate this argument, we plot in figure \ref{fig7_VelocityVsAw} the mean value of $v_y$ against $A\omega$ for $t\leq 1.5 sec$ for all the amplitudes and frequencies. The plot shows clearly that sinking occurs only when $A\omega \geq g_c=(120\pm 5)m/sec^2$; they all cross from rising to sinking at the point 
$(v_y, A\omega)=(0,g_c)$.
Using then the measured value of $g_c$ and approximating $\epsilon\approx 1/2$, we can use relation (\ref{SinkCondition}) to estimate the mean force required to fluidise a cavity base particle against the local friction at the initial level $h_0$: $\phi\left(h_0\right)\approx 3N$.

Another way to support this result is by plotting the contour line $g_c=120m/sec^2$ in the phase diagram, figure \ref{fig6_PhaseDiagramSettlingDepth} (dashed black line). This line follows closely the phase boundary between rising and sinking.
Equation (\ref{SinkCondition}) also shows that, unlike $f_c$ and $A_{min}$, the value of $g_c$ depends on features of both the IO, through its mass, and the medium, through $\phi$, which depends on the inter-particle friction and the depth $h$.

 \begin{figure}[!h]
\includegraphics[width=0.48\textwidth]{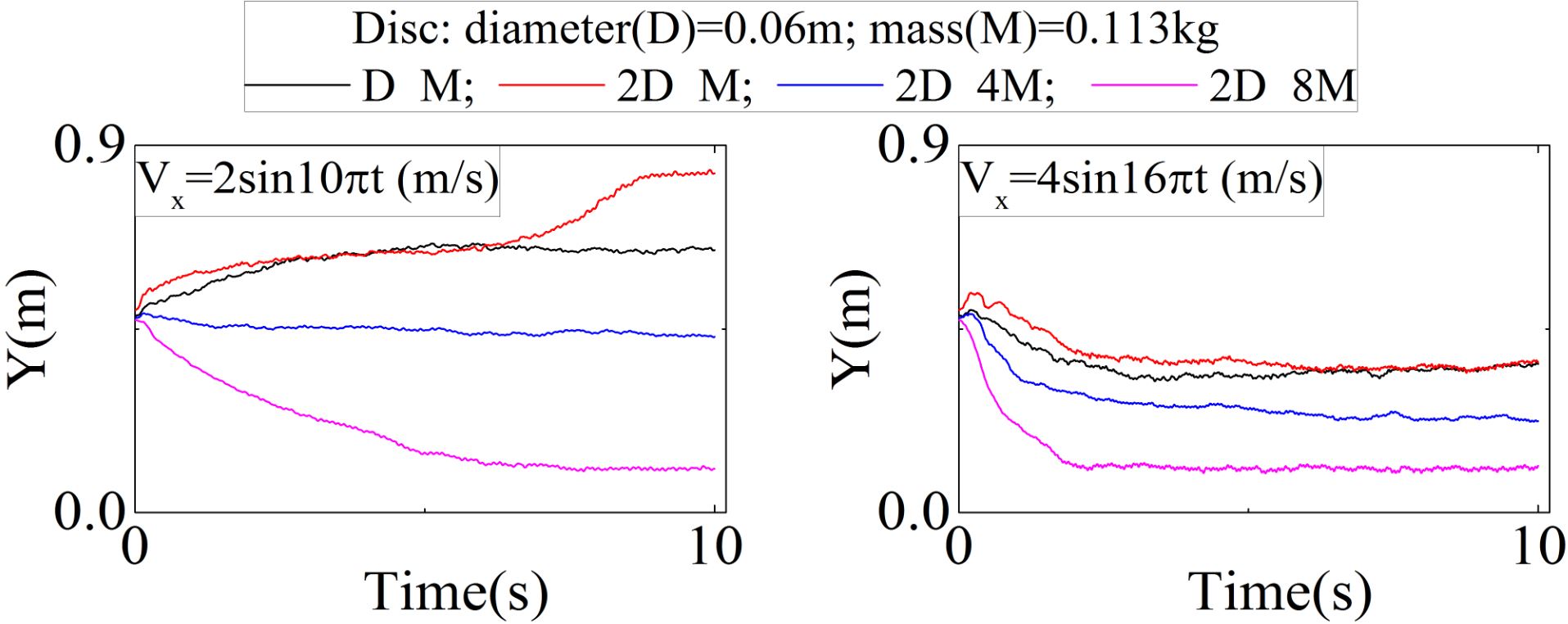}
\caption{The effects of the IO diameter and mass on the dynamics.}
\label{DMTest}
\end{figure}

Finally, to test our results, we ran simulations, with different IO properties: $D\to 2D$ and $M\to 2M, 4M, 8M$, for two sets of parameters: (i) $A=2$m/sec and $f=5$Hz and (ii) $A=4$m/sec and $f=8$Hz. The different depth evolutions are shown in figure \ref{DMTest} and provide strong support of the analysis. Increasing the diameter to $2D$ is predicted by relation (\ref{YRise0}) to reduce $f_c$ and indeed, in set (i), the IO rises (red line) above the original system (black line), for the same mass. In contrast, relation (\ref{SinkCondition}) predicts independence of the sink condition, $A\omega=g_c$ of $D$ and this is supported by case (ii), in which the systems are hardly distinguishable. 
Relation (\ref{SinkCondition}) also predicts that the sink condition is inversely proportional to the mass and, indeed, the dynamics in case (ii) show that the denser the IO the lower it sinks.
 
\section{\label{Conclusion}Discussion and conclusions} 

To conclude, we studied, numerically and analytically, the vertical dynamics of a large, self-energised immersed object (IO), oscillating horizontally in a granular bed. We showed that the IO may rise, sink or remain at its initial depth, depending on the amplitude and frequency of the oscillation. We also showed that the IO settles eventually at a certain depth, unless it reaches the surface of the granular medium. This behaviour leads to a rich phase diagram of the rise rate as a function of the oscillation amplitude and frequency.
To explain and predict this behaviour, we presented a number of theoretical considerations, based on a cavity model. The model allowed us to derive the following general explicit results. 
(i) An expression for the critical frequency, above which the IO cannot rise. This frequency depends only on the gravitational acceleration and the IO's size. 
(ii) An expression for the minimal velocity amplitude, required to initiate vertical rise, which also depends only on the gravitational acceleration and the IO's size.
(iii) An expression for the critical acceleration amplitude, only above which the IO can sink. This critical acceleration was found to depend on characteristics of both the IO and the granular medium.
These derivations gave quantitative predictions, all of which agreed very well with the simulation results. Further support of our analysis was provided by changing the IO diameter and mass, all of which agreed with the derived relations. 

These results are useful to understand the sorting dynamics in granular materials, composed of largely similar size particles and a small fraction of larger ones, which is relevant to technological applications. They also have direct relevance to understanding some of the principles used by `sand swimmers' - animals that burrow and move in sand \cite{Do96,Maetal09,Maetal11a}: they suggest that these animals can control the depth of their locomotion simply by varying the velocity and frequency of their body horizontal undulations.

This work can be extended in a number of directions.  These include simulating larger systems, longer times, and more frequencies and amplitudes in order to be able to fill the rise rate phase diagram more accurately. This would also allow us to construct a phase diagram of the settling depth, which should be related qualitatively, but not necessarily quantitatively, to that of the rise rates. 
To test the expressions derived in this paper, further simulations should be run with different material parameters, including variable density, size and shape of the IO, which our model, combined with research in the literature \cite{Maetal11b}, suggests should affect the rise rate.
Another obligatory direction is theoretical: to construct an explicit equation of motion for the IO's vertical rise or sink. This will also allow to predict the final depth, which the IO settles into
Finally, our setup is amenable to experimental testing and we are looking forward to seeing such experiments. In our group, we are working on the theoretical modelling, as well as on some of the numerical extensions.

This work was funded, in part, by the Fundamental Research Funds of NUDT, grant ZK16-03-01.


\begin{thebibliography}{99}

\bibitem{Soetal06} R. Soller, and S. A. Koehler, Phys. Rev. {\bf E 74}, 021305 (2006).
\bibitem{Guetal13} F. Guillard, Y. Forterre, and O. Pouliquen, Phys. Rev. Lett. {\bf 110}, 138303 (2013).
\bibitem{Guetal14} F. Guillard, Y. Forterre, and O. Pouliquen, Phys. Fluids {\bf 26}, 043301 (2014).
\bibitem{Guetal15} F. Guillard, Y. Forterre, and O. Pouliquen, Phys. Rev. {\bf E 91}, 022201 (2015).
\bibitem{Knetal96} J. B. Knight, E. E. Ehrichs, V. Y. Kuperman, J. K.Flint, H. M. Jaeger, and S. R. Nagel, Phys. Rev. {\bf E 54}, 5726 (1996).
\bibitem{Esetal10} P. Eshuis, D. van der Meer, M. Alam, H. J. van Gerner, K. van der Weele, and D. Lohse, Phys. Rev. Lett. {\bf 104}, 038001 (2010).
\bibitem{ShMu98} T. Shinbrot, and F. J. Muzzio, Phys. Rev. Lett. {\bf 81}, 4365 (1998).
\bibitem{Moetal01} M. E. M\"obius, B. E. Lauderdale, S. R. Nagel, and H. M. Jaeger, Nature {\bf 414}, 270 (2001).
\bibitem{Sh04} T. Shinbrot, Nature {\bf 429}, 352 (2004).
\bibitem{Chetal08} C. Y. Cheuk, D. J. White, and M. D. Bolton, J. Geotech. Geoenviron. Eng. {\bf 134}, 154 (2008).
\bibitem{Meetal96} G. Metcalfe, and M. Shattuck, Physica A: Stat. Mech. Appl. {\bf 233}, 709 (1996).
\bibitem{WoWh98} R. J. K. Wood, and D. W. Wheeler, Wear {\bf 220}, 95 (1998).
\bibitem{Lo04a} D. Lohse, R. Rauhe, R. Bergmann, and D. van der Meer, Nature {\bf 432}, 689 (2004).
\bibitem{Lo04b} D. Lohse, R. Bergmann, R. Mikkelsen, C. Zeilstra, D. van der Meer, M. Versluis, K. van der Weele, M. van der Hoef, and H. Kuipers, Phys. Rev. Lett. {\bf 93}, 198003 (2004).
\bibitem{KaDu07} K. H. Katsuragi, and D. J. Durian, Nature Phys. {\bf 3}, 420 (2007).
\bibitem{Cletal15} A. H. Clark, A. J. Petersen, L. Kondic and R. P. Behringer, Phys. Rev. Lett. {\bf 114}, 144502 (2015).
\bibitem{Do96} K. J. Dowling, NASA report, (1996).
\bibitem{Maetal09} R. D. Maladen, Y. Ding, C. Li, and D. I. Goldman, Science {\bf 325}, 314 (2009).
\bibitem{Maetal11a} R. D. Maladen, Y. Ding, P. B. Umbanhowar, and D. I. Goldman, Int. J. Rob. Res. {\bf 30}, 793 (2011).
\bibitem{Coetal11} D. J. Costantino, J. Bartell, K. Scheidler, and P. Schiffer, Phys. Rev. {\bf E 83}, 011305 (2011).
\bibitem{Peetal11} B. Percier, S. Manneville, J. N. McElwaine, S. W. Morris, and N. Taberlet, Phys. Rev. {\bf E 84}, 051302 (2011).
\bibitem{Waetal03} C. R. Wassgren, J. A. Cordova, R. Zenit, and A. Karion, Phys. Fluids {\bf 15}, 3318 (2003).
\bibitem{Chetal03} D. Chehata, R. Zenit, and C. R. Wassgren, Phys. Fluids {\bf 15}, 1622 (2003).
\bibitem{Dietal11} Y. Ding, N. Gravish, and D. I. Goldman, Phys. Rev. Lett. {\bf 106}, 028001 (2011).
\bibitem{Maetal11b} R. D. Maladen, P. B. Umbanhowar, Y. Ding, A. Masse, and D. I. Goldman, in proceedings of IEEE ICRA, pp. 1398 (2011).
\bibitem{PoDi13} F. Q. Potiguar, and Y. Ding, Phys. Rev. {\bf E 88}, 012204 (2013).
\bibitem{Shetal09} T. Shimada, D. Kadau, T. Shinbrot, and H. J. Herrmann, Phys. Rev. {\bf E 80}, 020301 (2009).
\bibitem{CoBo10} C. N. Cohen, and J. H. Boyle, Cont. Phys. {\bf 51}, 103 (2010).
\bibitem{Maetal11c} R. D. Maladen, Y. Ding, P. B. Umbanhowar, A. Kamor, and D. I. Goldman, J. Roy. Soc. Interface {\bf 8}, 1332 (2011).
\bibitem{Beetal96} N. V. Brilliantov, F. Spahn, J. M. Hertzsch and T. P\"oschel, Phys. Rev. {\bf E 53}, 5382 (1996).
\bibitem{TaMo82} Y. Tatara, and N. Moriwaki, Bull. JSME {\bf 25}, 631 (1982).
\bibitem{Scetal98} T. Schwager and T. P\"oschel, Phys. Rev. {\bf E 57}, 650 (1998).
\bibitem{Ge66} Gear C W. {\it The numerical integration of ordinary differential equations of various orders}, Argonne National Laboratory Report (1966).
\bibitem{Ge71} Gear C W. {\it Numerical initial value problems in ordinary differential equations}, (Prentice Hall PTR, NJ, USA 1971).

\end{thebibliography}
\end{document}